\documentclass[%
 aip,
 amsmath,amssymb,
preprint,%
]{revtex4-1}

\usepackage{graphicx}
\usepackage{dcolumn}
\usepackage{bm}

\usepackage[utf8]{inputenc}
\usepackage[T1]{fontenc}
\usepackage{mathptmx}
\usepackage{etoolbox}
\usepackage{upgreek}
\usepackage{setspace}
\makeatletter
\def\@email#1#2{%
 \endgroup
 \patchcmd{\titleblock@produce}
  {\frontmatter@RRAPformat}
  {\frontmatter@RRAPformat{\produce@RRAP{*#1\href{mailto:#2}{#2}}}\frontmatter@RRAPformat}
  {}{}
}%
\makeatother
\begin{document}

\preprint{AIP/123-QED}

\title[Cascaded forward Brillouin lasing in a chalcogenide whispering gallery mode microresonator]{Cascaded forward Brillouin lasing in a chalcogenide whispering gallery mode microresonator}
\author{Thariq Shanavas}
 \affiliation{Department of Physics, University of Colorado, Boulder, CO 80309, USA}
\author{Michael Grayson}
\affiliation{Department of Electrical, Computer and Energy Engineering, \\University of Colorado, Boulder, CO 80309, USA}
\author{Bo Xu}
\affiliation{Department of Physics, University of Colorado, Boulder, CO 80309, USA}
\author{Mo Zohrabi}
\affiliation{Department of Electrical, Computer and Energy Engineering, \\University of Colorado, Boulder, CO 80309, USA}
\author{Wounjhang Park}
\affiliation{Department of Electrical, Computer and Energy Engineering, \\University of Colorado, Boulder, CO 80309, USA}
\affiliation{Materials Science Engineering Program, \\University of Colorado, Boulder, CO 80309, USA}
\author{Juliet T. Gopinath}
\affiliation{Department of Physics, University of Colorado, Boulder, CO 80309, USA}
\affiliation{Department of Electrical, Computer and Energy Engineering, \\University of Colorado, Boulder, CO 80309, USA}
\affiliation{Materials Science Engineering Program, \\University of Colorado, Boulder, CO 80309, USA}
\email{julietg@colorado.edu}
\date{\today}

\begin{abstract}
We report the first observation of cascaded forward stimulated Brillouin scattering in a microresonator platform. We have demonstrated 25 orders of intramodal Stokes beams separated by a Brillouin shift of 34.5 MHz at a sub-milliwatt threshold at 1550 nm. An As$_2$S$_3$ microsphere of diameter 125 $\upmu$m with quality factor $1\times 10^6$ was used for this demonstration. Theoretical modeling is used to support our experimental observations of Brillouin shift and threshold power. We expect our work will advance the field of forward stimulated Brillouin scattering in integrated photonics with applications in gas sensing and photonic radio frequency sources.
\end{abstract}

\maketitle

\section{Introduction}
Chalcogenide glasses have been recognized in the past decade as a candidate for low-threshold nonlinear optics because of their transparency in the infrared region, low glass transition temperature, and large nonlinearities\cite{petit2009compositional, eggleton2011chalcogenide, zakery2003optical, scheer2011chalcogenide, adam2014chalcogenide}. They also exhibit many desirable photoinduced phenomena including photocrystallization \cite{lyubin1997polarization}, photopolymerization \cite{fritzsche1993origin}, photodecomposition \cite{owen1985photo}, photocontraction or expansion \cite{chopra1981origin}, photovaporization \cite{owen1985photo}, and photodissolution of metals such as silver \cite{owen1985photo,seddon1995chalcogenide}. These changes allow us to modify the optical characteristics of chalcogenide glasses such as electronic band gap, refractive index, and optical absorption coefficient to fabricate a wide variety of optical devices. Additionally, the low glass transition temperature of chalcogenide glasses allows fabricated devices to be reflowed to produce ultrasmooth optical surfaces \cite{bae2020chip,hu2010optical,zhao2019exploration}.

Stimulated Brillouin scattering (SBS) is one of the strongest third-order nonlinearities in most optical media. SBS manifests through the transfer of energy from the optical pump beam to an optical Stokes beam, either in the same or opposite direction as the pump. SBS is used in narrow-linewidth lasers \cite{loh2019ultra,loh2015dual}, slow light \cite{zadok2011stimulated}, optical cooling \cite{bahl2012observation}, optical isolators \cite{ma2020chip}, and distributed sensing \cite{horiguchi1995development}. The conservation of energy and momentum requires that $\omega_a=\omega_p-\omega_s$ and $q=k_p\pm k_s$ where \(\omega_a\), \(\omega_p\) and \(\omega_s\) are the frequencies and $q$, \(k_p\) and \(k_s\) are the wavenumbers of the acoustic, pump and Stokes waves respectively. The sign in the momentum matching condition is positive (i.e., wavenumbers add up) for backward stimulated Brillouin scattering (BSBS) and negative for forward stimulated Brillouin scattering (FSBS). Since the pump and Stokes beams typically differ by no more than a few GHz in frequency, they are approximately equal in wavenumber, i.e. \(k_p\approx k_s\). Therefore, FSBS is mediated by phonons of low wavenumber, \(q\approx 0\), while BSBS is mediated by phonons of higher wavenumber, \(q\approx 2k_p\). Even though FSBS is understood to be theoretically a stronger nonlinearity than BSBS with all else being equal \cite{qiu2013stimulated}, many common resonator geometries do not have resonant acoustic modes for phonons with low wavenumber. Notably, fiber ring resonators do not support strong FSBS although they are widely used for BSBS lasers \cite{hill1976cw,wang2021tailorable}. This is because step index fibers do not have acoustic eigenmodes for low-momentum phonons.

A resonator that satisfies phase-matching conditions for the optical pump, optical Stokes, and acoustic phonon waves exhibits SBS at sub-mW thresholds\cite{mirnaziry2017lasing} due to the tight confinement of the optical fields. To exploit the many advantages of nonlinear optics using SBS in a small form factor, whispering gallery mode resonators are often used \cite{lin2016opto,del2013laser}. The resonance condition for the Stokes beam can be met by matching the free spectral range (FSR) of the resonator to the SBS shift so that the pump and Stokes beams can be excited to adjacent resonances with nearby azimuthal mode orders. Such resonators with FSR matching may support cascaded SBS. In fact, cascaded BSBS with GHz-scale shifts has been demonstrated in resonators of mm-scale diameter \cite{lin2016opto,del2013laser,grudinin2009brillouin}. The FSR, unfortunately, scales inversely with the length of the resonator, which makes it difficult to fabricate a high-quality resonator that matches typical FSBS MHz shifts. For this reason, FSBS in a resonator has only been demonstrated between optical modes with different radial and azimuthal mode orders (i.e. intermodal FSBS) \cite{yu2022investigation,bahl2011stimulated}. This technique has the downside that it would be difficult to consistently reproduce the same Brillouin shift across devices, as the frequency difference between modes without at least one common mode order (radial or azimuthal) is highly sensitive to device geometry, placing a strict limit on fabrication tolerance. It is also not possible to demonstrate cascaded SBS using this technique since higher-order modes with differing radial and azimuthal mode orders are not equally spaced in frequency. For instance, in the first report of FSBS in a microresonator Bahl et al.\cite{bahl2011stimulated} reports forward Brillouin scattering between a pump mode (640, 2, 1) and Stokes mode (630, 1, 3) with a Brillouin shift of 57.8 MHz where the numbers in brackets are respectively azimuthal, radial and polar mode orders. Since there is no optical mode 57.8 MHz below the first Stokes mode (630, 1, 3), a cascaded process is not observed.

In this work, we avoid the technical difficulty of matching the FSR to the Brillouin shift by using intramodal Brillouin scattering (i.e., the Stokes beam is scattered within the same optical resonance as the pump beam). This is enabled by a chalcogenide glass microresonator platform with a forward Brillouin shift that is smaller than the width of the optical resonance. We rely on finite element modeling to choose a microresonator geometry that supports low-wavenumber resonant acoustic modes required to meet the phase matching condition. Using this approach, we demonstrate up to 25 orders of cascaded intramodal FSBS at a sub-mW threshold. The measured threshold of 0.91~mW is comparable to a previous report of  5 orders of cascaded BSBS in a silica microsphere at 0.6~mW \cite{guo2015ultralow}.

In addition to the applications of Brillouin lasers in narrow-linewidth optical sources or slow-light generation, cascaded SBS can be used to improve optical sensors and photonic frequency synthesizers. The Brillouin frequency shift is cumulative over cascaded Stokes beams. Therefore, a small change in Brillouin shift leads to higher order Stokes beams experiencing a greater net change in frequency than lower Stokes orders. This effect can be leveraged to improve the sensitivity of Brillouin-based gas sensors that rely on a change in the Brillouin shift as a function of gas concentration \cite{yao2017graphene}. For example, an increase in Brillouin shift of $\delta f$ leads to the frequency of the 25th Stokes beam decreasing by $25\times\delta f$, improving the sensitivity of the detector 25-fold.

Cascaded SBS on chip has been adopted to build photonic frequency synthesizers to produce low-noise radio frequency signals\cite{fortier2011generation,gundavarapu2019sub,li2013microwave}. The technique presented in this paper does not require the free spectral range of the resonators to be matched to the Brillouin shift, therefore we expect this work to loosen the fabrication tolerances for microresonators used in photonic frequency synthesizers. Additionally, since the forward Brillouin frequency shift is strongly dependent on the geometry of the resonator \cite{wang2021tailorable}, we expect this result to advance the state of the art in low-noise tunable photonic frequency synthesizers.

\section{Microresonator fabrication and modeling}
High-quality chalcogenide microspheres were fabricated using commercially available arsenic sulfide (As$_2$S$_3$) fiber (IRFlex -- IRF-S-6.5). First, the fiber was tapered down to $\sim$ 20 \textmu m by heating the fiber to the glass transition temperature using a resistive heater and pulling on both sides using a motorized stage. The taper was then broken and the tapered end was molten by bringing it close to a red-hot metal plate. The surface tension caused the molten glass to form a sphere, which was then allowed to cool down. It is possible to produce microsphere resonators of Q-factor over \(10^7\) using this technique. This is close to the theoretical upper limit due to absorption \cite{yu2022investigation,zhu2019nonlinear}.

The optical and acoustic modes of the resonator were modeled using the commercial COMSOL multiphysics package\cite{multiphysics1998introduction}. Simulations were performed for intramodal SBS, where the pump mode and all Stokes orders have the same azimuthal, polar and radial mode numbers. To satisfy this condition, the width of the resonance needs to be at least 300~MHz when the resonator is coupled to a tapered fiber (i.e. the loaded Q factor has to be less than $6.6\times 10^5$). Therefore, measurements and simulations were performed in the overcoupled regime. This is in sharp contrast to the common practice of using the highest quality factor resonator that is possible to fabricate and undercoupling it for backward Brillouin lasers.

Since the resonances were all at least 300~MHz wide, several Stokes orders that were separated by a few tens of MHz would fit within the same resonance. The phase-matching condition requires that the azimuthal mode of the acoustic wave is the difference in mode numbers of the pump and Stokes beams. For FSBS, we hence have a standing acoustic wave with an azimuthal mode number equal to zero.
Using the electric and displacement fields from the resonant optical and acoustic modes respectively, we calculated the SBS gain in the resonator. The SBS gain of a single acoustic mode has a Lorentzian shape and a peak value of \cite{qiu2013stimulated}, 
\begin{equation}
    \Gamma = \frac{2\omega Q_m}{\Omega^2\nu_{gp}\nu_{gs}}  \frac{|\langle\mathbf{f,u_m}\rangle|^2}{\langle\mathbf{E_p},\epsilon\mathbf{E_p}\rangle\langle\mathbf{E_s},\epsilon\mathbf{E_s}\rangle\langle\mathbf{u_m},\rho\mathbf{u_m}\rangle}
    \label{eq:gain}
\end{equation}
where \(\omega = \omega_p\approx\omega_s\) is the frequency of the optical beams, $\Omega$ is the frequency of acoustic wave or the SBS shift, \(\nu_{gp}\) is the group velocity of the pump, \(\nu_{gs}\) is the group velocity  of the Stokes beam, \(\epsilon\) is the permittivity of the material, \(\rho\) is the density, \(\mathbf{E_p}\) and \(\mathbf{E_s}\) are the unnormalized electric fields from the pump and Stokes beams respectively, \(\mathbf{u_m}\) is the unnormalized displacement vector from the resonant acoustic wave, and \(\mathbf{f}\) is the net force due to electrostriction and radiation pressure. The angled brackets indicate surface integrals performed on a radial plane perpendicular to the direction of propagation of the optical beams.

The phonon quality factor \(Q_m\) is defined as twice the ratio of phonon frequency over the damping rate. Following the experimental results of Bahl et al. \cite{bahl2011stimulated}, we assume that the phonon quality factor for an Arsenic Sulphide microresonator is limited by the phonon damping rate from the material. We estimate the damping rate as \(\pi\Delta\nu_B\)\cite{debut2000linewidth}, where \(\Delta\nu_B\) is the gain bandwidth and obtain \(Q_m = 467\) from the experimental results of Pant et al.\cite{pant2011chip}. We also performed finite element simulations to include the phonon loss rate through the stem of the microresonator using COMSOL. Our simulations yielded \(Q_m\) to be about 800. Since \(Q_m\) is difficult to measure experimentally and the simulations are strongly dependent on errors in reported material values, we use an order of magnitude approximation and take \(Q_m = 500\) for our calculations.

Electrostriction force is a deforming force experienced by all dielectrics in the presence of non-uniform electric fields. The highly localized whispering gallery mode resonances in optical microresonators hence induce electrostriction forces in the bulk of the resonator. The electrostriction force is derived from the gradient of the electrostrictive tensor given by \cite{qiu2013stimulated},
\begin{equation}
    \sigma_{ij}=-\frac{1}{2}\epsilon_0n^4p_{ijkl}E_kE_l
\end{equation}
where \(n\) is the refractive index, \(p_{ijkl}\) is the photoelastic tensor, and \(E_k\) and \(E_l\) are the net electric fields in the direction denoted by indices \(k\) and \(l\). 

The radiation pressure on the other hand is the pressure exerted on a dielectric interface due to the exchange of momentum between the dielectric and an electromagnetic field incident on its surface. The radiation pressure is derived from the Maxwell stress tensor. As the radiation pressure acts perpendicularly to the dielectric interface \cite{qiu2013stimulated} and the acoustic eigenmode corresponding to the Brillouin shift oscillates parallel to the interface, the radiation pressure has no contribution to the Brillouin gain. Hence, we excluded radiation pressure from our calculations to reduce computational complexity.

\begin{figure}
    \centering
    \includegraphics[width=\textwidth]{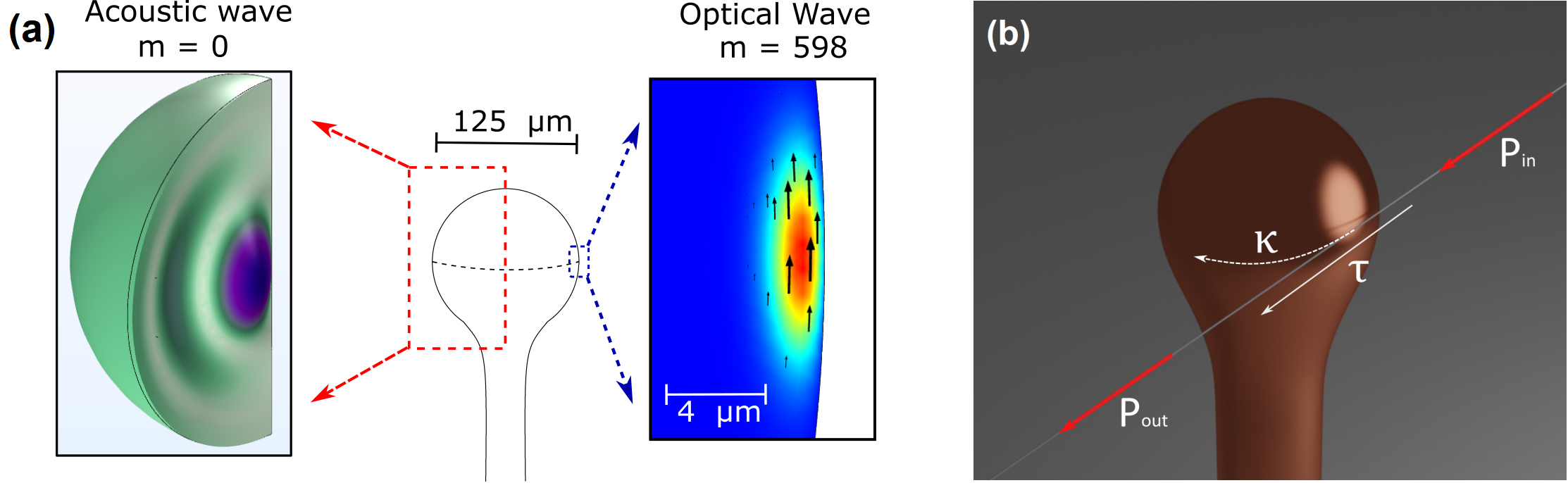}

    \caption{(a) Simulations of optical and acoustic modes for phase-matched Brillouin scattering in an As$_2$S$_3$ microsphere of diameter 125 $\upmu$m. The acoustic eigenmode oscillating at 33.2 MHz is shown on the left. Purple highlights the region with the highest deformation. Deformation is exaggerated to show the effect. The electric fields to the right correspond to optical excitation at 1550 nm with the azimuthal mode order represented by $m$. The colors show the strength of the electric field, and the arrows show the electric field direction. (b) Rendering of the tapered fiber coupling. $\kappa$ and $\tau$ are the coupling constants normalized to $\sqrt{\kappa^2+\tau^2}=1$.}
    \label{fig:simulation}
\end{figure}

When the pump power is below the Brillouin lasing threshold, pump photons scatter off thermally populated phonons to generate a very small amount of Stokes power. In this regime, we may use the small signal approximation (SSA) that the Stokes power is much lower than the pump power. We also ignore pump depletion and nonlinear losses. When the SSA is valid, the Stokes power coupled out of the resonator $P_s^{out}$ is related to the input Stokes power $P_s^{in}$ (which arises from thermal phonons) as \cite{mirnaziry2017lasing},

\begin{equation}
    P_s^{out} = \left| \frac{\left|\tau\right|-G}{1-\left|\tau\right|G}  \right|^2 P_s^{in}
    \label{eq:StokesAmplification}
\end{equation}
where $\tau$ is the coupling constant [Fig.~\ref{fig:simulation}(b)] and $G$ is the round-trip envelope gain given by \cite{mirnaziry2017lasing},

\begin{equation}
    G = \exp\left[ -\frac{\alpha L}{2} + \frac{\Gamma}{2 \alpha} P_p^{in}(1-e^{-\alpha L})\right]
\end{equation}

Here, $P_p^{in}$ is the input pump power, $\alpha$ is the total loss and $L$ is the length of the cavity. To account for the various losses including material, scattering and radiation losses, \(\alpha\) is estimated from the measured Q factor as \(\alpha = 2\pi n/Q\lambda\) \cite{gorodetsky1996ultimate}.  At the Brillouin lasing threshold, the SSA breaks down as $G$ approaches $1/\left|\tau\right|$ and Eq. \ref{eq:StokesAmplification} goes to infinity. The pump power where the SSA breaks down is therefore taken to be the lasing threshold \cite{mirnaziry2017lasing}. While fitting the SSA model in Eq. \ref{eq:StokesAmplification} to the experimental data, $P_s^{in}$ only introduces a scaling factor in the order of $10^{-11}$ if we assume only a single photon is generated per second per unit frequency by quantum fluctuations and thermal phonons ($P_s^{in} = v_g h f_s \approx 10^{-11} $, $v_g$ is the group velocity, $h$ is the Planck's constant and $f_s$ is the Stokes frequency \cite{mirnaziry2017lasing}). Since $\alpha$ is the loss and $\Gamma$ is obtained from simulations of the acoustic and optical eigenmodes of the system [Eq. \ref{eq:gain}], the only remaining free parameter in the model is $\tau$, the coupling constant. We extract $\tau$ from the experimental data to estimate the lasing threshold using the SSA model in Section \ref{sec:measurement}.

\section{Optical Measurement}
\label{sec:measurement}

\begin{figure}
     \centering
     \includegraphics[height = 6cm]{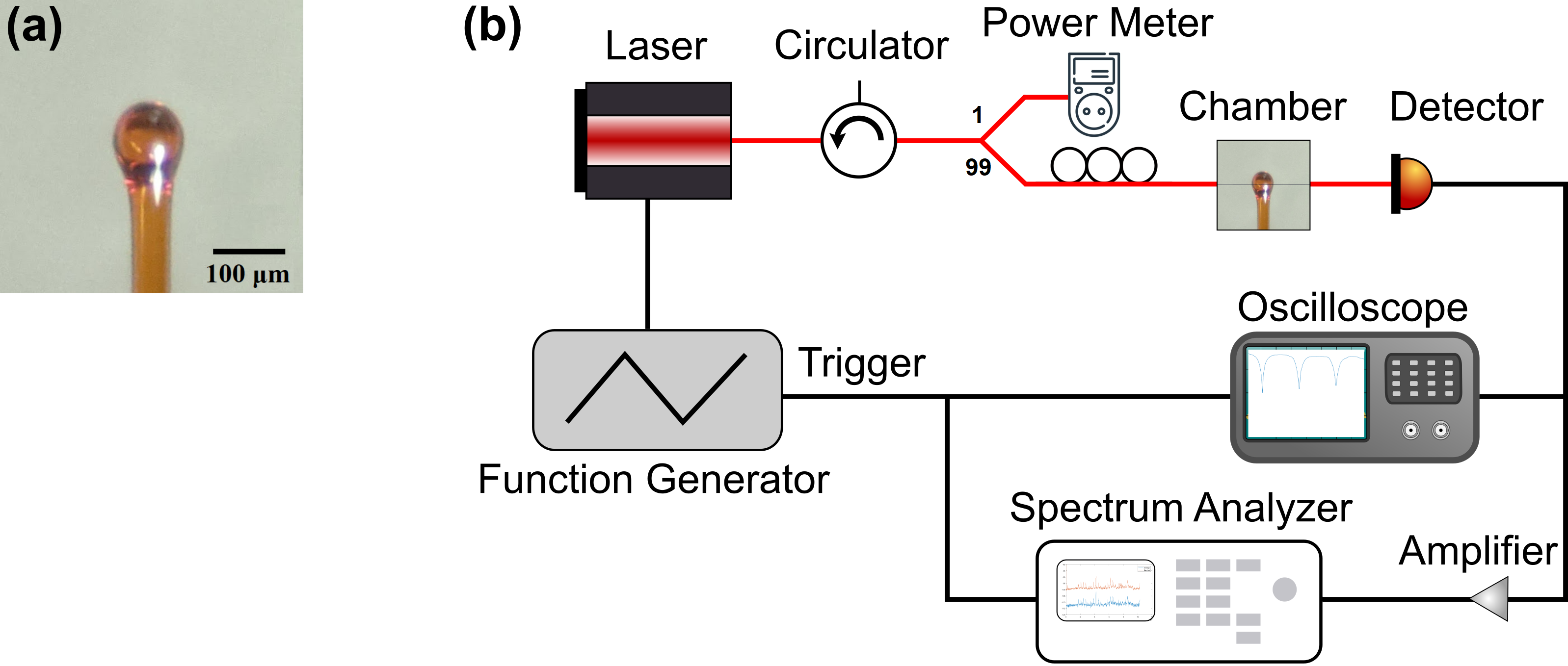}
        \caption{(a) Optical microresonator of diameter 100 $\upmu$m fabricated from tapered As$_2$S$_3$ fiber. (b) The experimental setup consists of a tunable diode laser at 1550 nm that was coupled into a tapered silica fiber. A 99:1 splitter was used to monitor the pump power. The resonator was brought close to the taper via a 3-axis stage with a sub-micron resolution. A function generator provides a triangle wave signal to the laser to sweep the frequency within a few tens of GHz. The circulator prevents back reflections into the diode laser.}
        \label{fig:setup}
\end{figure}

\begin{figure}
    \centering
    \includegraphics[width=0.95\textwidth]{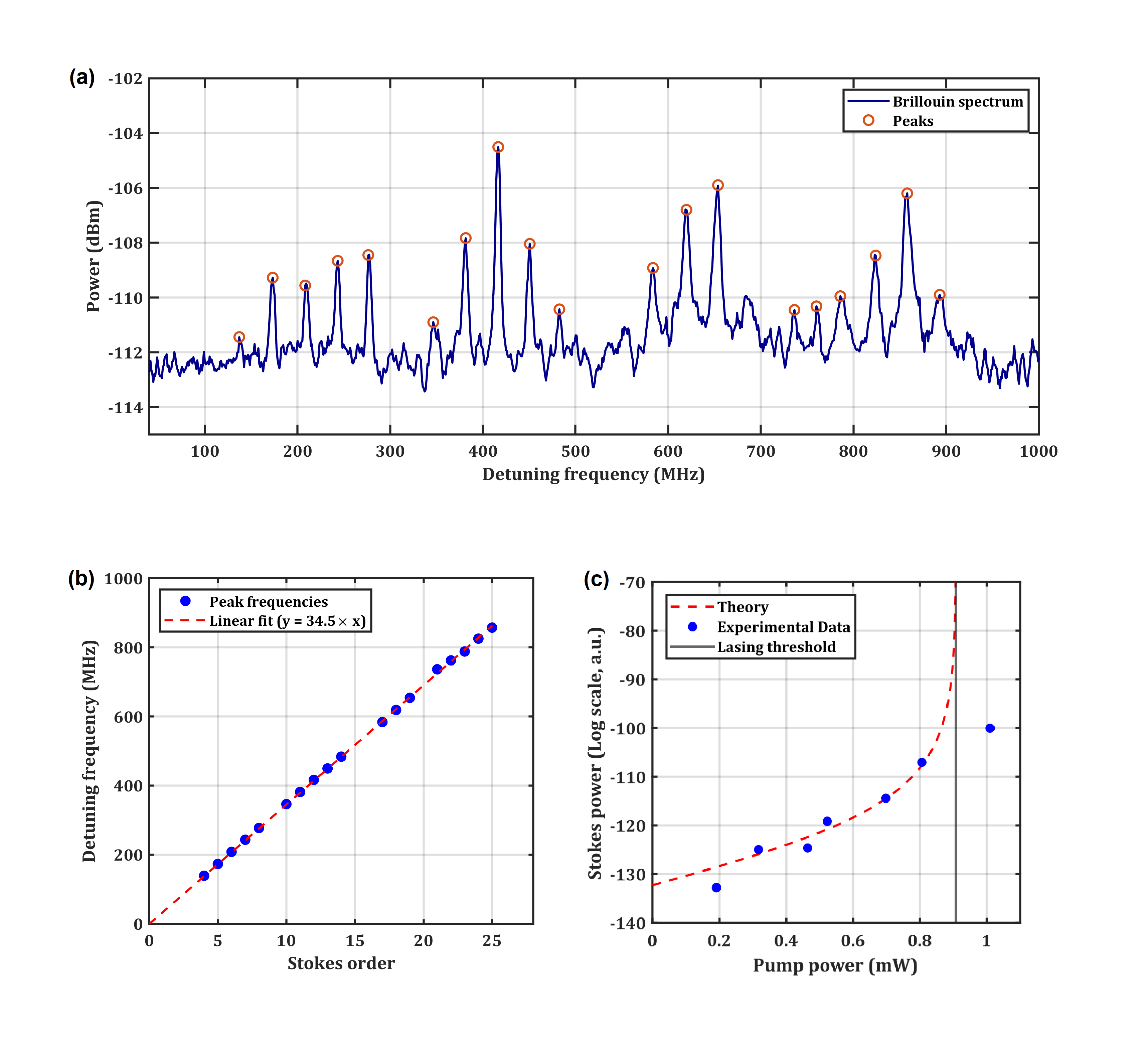}
    \caption{(a) Cascaded FSBS spectrum in a 125 $\upmu$m sphere with an unloaded quality factor $1 \times 10^6$. The x-axis is the detuning from the pump and the y-axis on the left shows the power of the beat note. This measurement was taken in the overcoupled regime using a 1.2 GHz bandwidth detector and the spectrum was averaged over 41 sweeps. The spectral lines were broadened because of the variation in acoustic eigenfrequency across sweeps due to thermal effects. The powers of the spectral lines are uneven since the amplifier circuit had parasitic absorptions at 105(2n+1) MHz (n = 0, 1, 2 ...). The red circles highlight the location of the peaks in the spectrum. (b) The frequency shifts of the Stokes beams from the pump, which are observed to be integer multiples of 34.5 MHz. The acoustic mode simulated in Fig.~\ref{fig:simulation}(a) agrees well with the observed FSBS shift. (c) The power of the first Stokes beam vs. pump power for a 110~$\upmu$m resonator of Q factor $2.2\times 10^6$ showing an FSBS shift at 18.3 MHz. The experimental data is fit to the small signal approximation (SSA) model \cite{mirnaziry2017lasing}. The SSA model is valid only below the Brillouin lasing threshold - therefore, the one data point above the threshold was not fit to the model. The SSA model breaks down at the threshold shown by the vertical black line, and goes to infinity. The stimulated lasing threshold was found to be 910 $\pm$ 20 $\upmu$W, beyond which cascaded Stokes beams were observed.}
    \label{fig:data125u}
\end{figure}

\begin{figure}
    \centering
    \includegraphics[width=\textwidth]{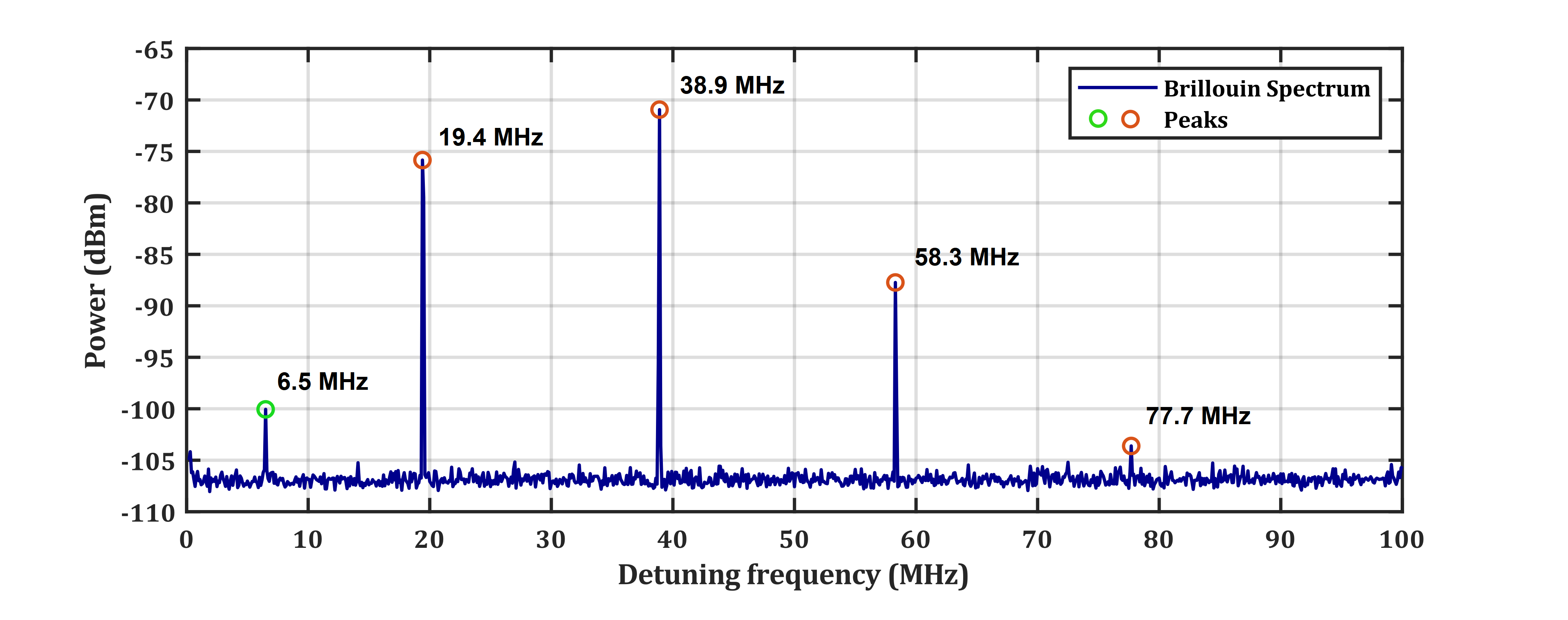}
    \caption{Cascaded FSBS spectrum in a 100 $\upmu$m sphere with unloaded quality factor 6.2~$\times~10^5$. We did not use an amplifier for this measurement and time averaging was not necessary. The spectrum was taken using a detector of high responsivity and 40 MHz rated bandwidth. The peaks highlighted in red are part of the cascaded spectrum, with a Brillouin shift of 19.4 MHz. The 6.5 MHz peak results from an acoustic mode having a weaker overlap with the pump and therefore does not cascade. The pump was blue-detuned from the resonance, hence the 19.4 MHz peak experiences more loss than the 38.9 MHz peak. The number of Stokes orders observed is likely bandwidth-limited by the detector.}
    \label{fig:data100u}
\end{figure}

The experimental setup is shown in Fig.~\ref{fig:setup}(b). An SMF-28 silica fiber was tapered to \textasciitilde~500~nm radius by heating the fiber to above the glass transition temperature using a butane flame while stretching the fiber from both ends using motorized stages. The laser source was a single frequency mode-hop-free tunable CW laser in a 1550 nm band (Toptica CTL 1550). The resonator was mounted on a 3-axis piezo-actuated stage (Thorlabs MAX312D) with 20~nm resolution and brought within 100~nm of the tapered fiber. The system was imaged using a microscope with a long-distance objective (Mitutoyo 10X ICO 0.28 NA, 33.5 mm working distance) from the top. The system was enclosed in a chamber to minimize disturbances due to temperature fluctuations and air currents. The transmitted light was measured using an InGaAs photodetector of 1.2 GHz bandwidth (Thorlabs DET01CFC) for the spectrum in Fig.~\ref{fig:data125u}(a) and an InGaAs photodetector of 40~MHz bandwidth (Thorlabs DET10C) for the spectrum in Fig.~\ref{fig:data100u}. A 20~dB amplifier was used with the high-bandwidth detector to make up for the lower responsivity. The signal was recorded using an oscilloscope and an RF spectrum analyzer (Keysight N9030B). A function generator provided a triangle wave signal to the laser to sweep the frequency by a few tens of GHz, and a trigger signal to the RF spectrum analyzer and the oscilloscope. The laser was tuned into resonance by scanning for the characteristic dip in transmission across the tapered fiber on the oscilloscope. This dip is measured while operating at undercoupled condition to measure the intrinsic Q factor of our resonators. The Stokes beams in the resonator are coupled back into the tapered fiber. The electrical spectrum analyzer was used to pick up beat notes between the Stokes beams and the pump that are typically too close for an optical spectrum analyzer to resolve. Using this setup, we observed the cascaded Brillouin spectrum shown in Fig.~\ref{fig:data125u}(a) from a 125 $\upmu$m resonator of unloaded quality factor $1\times 10^6$ and the spectrum in Fig.~\ref{fig:data100u} from a 100~$\upmu$m resonator [Fig.~\ref{fig:setup}(a)] of unloaded quality factor $6.2\times 10^5$.

Fig.~\ref{fig:data125u}(b) shows a plot of peak frequencies from Fig.~\ref{fig:data125u}(a) as a function of Stokes order. The peak frequencies were found to lie at integer multiples of 34.5~MHz, with the highest detuning observed at 25 times the Brillouin shift (i.e., 25 Stokes orders). This is consistent with our simulations [Fig.~\ref{fig:simulation}(a)] which show a resonant acoustic wave at 33.2~MHz. The peak frequencies in Fig.~\ref{fig:data100u} were similarly found to lie at integer multiples of 19.4~MHz. This is also consistent with our simulations for a 100 $\upmu$m resonator showing resonant acoustic waves at 17.4 MHz. The mismatch in frequencies (34.5 MHz in experiment vs 33.2 MHz in simulation, 19.4 MHz in experiment vs. 17.4 MHz in simulation) is attributed to the deviation in Young's modulus of the commercial As$_2$S$_3$ fiber from the reported value of bulk material \cite{black1957properties} and the perturbation in mode shape due to the stem of the resonator which was not included in the simulation.

\begin{figure}
    \centering
    \includegraphics[width=0.5\textwidth]{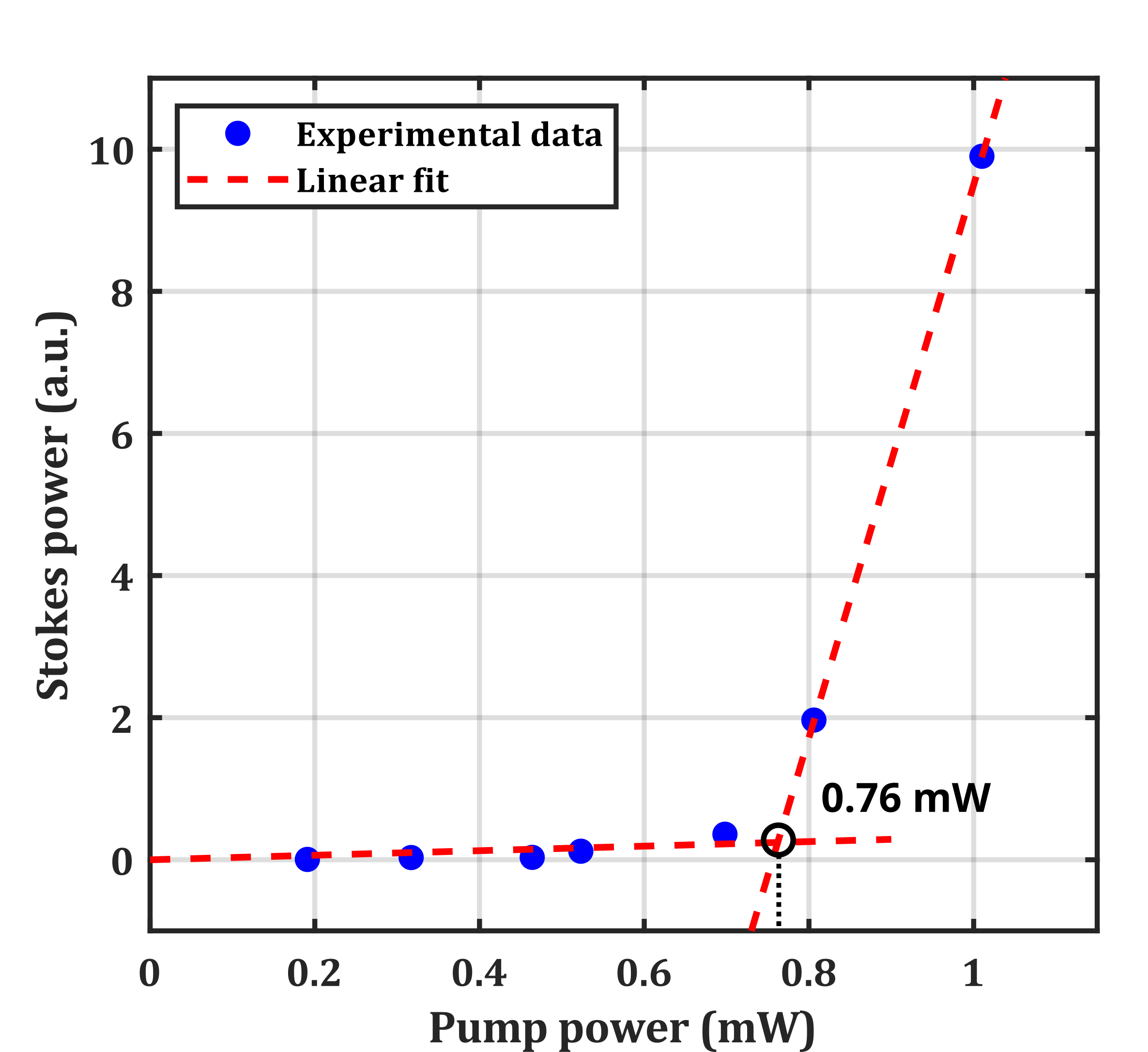}
    \caption{The data in Fig \ref{fig:data125u}(c), presented in a linear scale. The knee of the Stokes power output indicates the lasing threshold is reached at about 0.76 mW. This is slightly lower than the prediction from fitting the small-signal model at 0.91 mW}
    \label{fig:threshold_linear}
\end{figure}

To estimate the lasing threshold, we used a microsphere of diameter 110~$\upmu$m with quality factor $2.2 \times 10^6$, showing cascaded Stokes beams separated by 18.3~MHz. As the pump power was reduced, the first Stokes line was the last to drop below the noise floor of the spectrum analyzer. We then slowly ramped up the pump power and recorded the power of the first Stokes line until the resonator suffered a degradation in Q factor.

Our experimental setup performs heterodyne measurement between the pump and Stokes beams to achieve high frequency resolution. But since the electrical spectrum analyzer was used to pick up beat notes between the pump and Stokes beams, we need to convert the measured signal $S$ in dB to optical power units for measuring the lasing threshold. The detector current from superposing optical beams $I_1$ of frequency $f_1$ and $I_2$ of frequency $f_2$, for detector sensitivity $R$ is \cite{razdan2002demonstrating},

\begin{equation}
    I(t) = R\left[ I_1+I_2+2\sqrt{I_1I_2}\cos(2\pi(f_1-f_2)t)\right]
\end{equation}

Since the spectrum analyzer records the AC component of the signal, and taking $I_1$ to be the pump and $I_2$ as the Stokes and using subscripts $p$ for pump and $s$ for Stokes,

\begin{equation}
    S=10\log_{10} \left(\frac{2R\sqrt{I_pI_b}}{I_{ref}}\right) (dB)
\end{equation}
where $S$ is the measured signal and $I_{ref}$ is the internal reference current of the analyzer. To obtain Stokes power as a function of varying pump power given measured value of $S$ vs. $I_p$,
\begin{equation}
    I_b = \frac{1}{I_p}\left(\frac{1}{2R} I_{ref}10^{S/10}\right)^2
\end{equation} 

The power of the first Stokes line as a function of pump power is shown in Fig.~\ref{fig:data125u}(c). Finite element simulations of this sphere revealed a resonant acoustic mode at 16.1 MHz that satisfies the phase matching condition, close to the observed FSBS shift at 18.3~MHz. Using Eq. \ref{eq:gain}, we estimated the Brillouin gain $\Gamma$ of the resonator to be $2.57~\times~10^5$~m$^{-1}$W$^{-1}$. The calculated value of $\Gamma$ was used to fit the theoretical model in Eq. \ref{eq:StokesAmplification} to the experimental data to obtain the coupling constant $\tau$ and the threshold power. The small signal approximation breaks down at the Brillouin lasing regime and the theoretical model goes to infinity at the threshold \cite{mirnaziry2017lasing}. From the fit, we found the Brillouin lasing threshold to be at 910 $\pm$ 20 $\upmu$W, for coupling constant $\tau = 0.92$. When the first order Stokes power is plotted against pump power on a linear scale in Fig. \ref{fig:threshold_linear}, a knee is observed at around 0.76 mW, confirming a sub-mW lasing threshold. The small disagreement in threshold power measurement from the different methods (0.91 mW vs 0.76 mW) is attributed to the ambiguity in the knee method\cite{newportknee}. As the quality factor degrades beyond about 1 mW pump power, we cannot take measurements in the milli Watt regime.

\section{Results and Discussion}
\label{sec:discussion}

We fabricated high-quality chalcogenide microresonators to study nonlinear effects in the transmitted light through a coupled tapered fiber. We have observed beat notes between downconverted Stokes beams and the pump beam in an RF spectrum analyzer when the narrow-linewidth tunable laser at 1550 nm was tuned into resonance in a 125 $\upmu$m sphere with quality factor $1\times10^6$. The beat notes were observed at integer multiples of 34.5 MHz, up to 25 orders which suggests a cascaded nonlinear downconversion process at resonance [Fig.~\ref{fig:data125u}(a)]. The observation is consistent with the simulation which suggested that this was Brillouin scattering mediated by a resonant acoustic mode. The slight mismatch in frequency (33.2 MHz in simulation as compared to 34.5 MHz in the experiment) was attributed to the deviation in Young's modulus of the commercial As$_2$S$_3$ fiber from the reported value of bulk material\cite{black1957properties} and the perturbation in mode shape from the stem of the microsphere which was not included in the simulation. We also observed a cascaded Brillouin spectrum in a smaller resonator of diameter 100 $\upmu$m with quality factor $6.2\times10^5$ (Fig.~\ref{fig:data100u}). The Brillouin shift was observed to be 19.4 MHz, close to its predicted value of 17.4 MHz from the simulation. 

Since the amplifier raised the noise floor of the spectrum in Fig.~\ref{fig:data125u}(a), the spectrum was averaged over 41 sweeps to increase the visibility of the peaks. However, the spectral lines were broadened to between \textasciitilde~3~MHz and \textasciitilde~9~MHz because of the variation in acoustic eigenfrequency across sweeps due to thermal fluctuations. The power of the spectral lines were uneven since the amplifier circuit had parasitic absorptions at 105(2n+1) MHz (n = 0, 1, 2 ...). When the gain of the amplifier (nominally 20 dB) fell below 16 dB around 105~MHz, 315~MHz, 525~MHz and 735~MHz, the signal was buried in the noise floor of the spectrum analyzer.

In contrast, the spectrum in Fig.~\ref{fig:data100u} was taken in one shot (i.e. no time averaging) and without using an amplifier. An amplifier was not required for the spectrum in Fig.~\ref{fig:data100u} on account of not coupling the resonator to the tapered fiber as strongly as we did for the resonator in Fig.~\ref{fig:data125u}(a). This is because we attempted to measure a GHz bandwidth comb with Fig.~\ref{fig:data125u}(a), which required the width of the resonance to be at the very least several GHz wide. This required that we had to overcouple the resonator until the loaded Q factor fell below $5~\times~10^4$. On the other hand, we only needed the resonance for the measurement in Fig.~\ref{fig:data100u} to be greater than a few 100~MHz wide. This condition was already met since the unloaded Q factor for the sphere used in Fig.~\ref{fig:data100u} was $6.2~\times~10^5$. As the Brillouin gain scales inversely with approximately the square of the loaded Q factor, the resonator used in Fig.~\ref{fig:data100u} had much stronger Stokes lines. We could thus obtain a high signal to noise ratio in Fig.~\ref{fig:data100u} without using an amplifier or averaging across multiple measurements.

We performed threshold measurements in a 110 $\upmu$m resonator with Q factor $2.2\times 10^6$ showing spectral lines shifted from the pump by 18.3 MHz [Fig.~\ref{fig:data125u}(c)]. Finite element simulations of this microsphere revealed a resonant acoustic mode at 16.1~MHz, close to the observed FSBS shift. We theoretically calculated the forward Brillouin gain $\Gamma$ to be  $2.57 \times 10^5$ m$^{-1}$W$^{-1}$ from Eq. \ref{eq:gain}. We note that this is much higher than typical BSBS gain, as the gain is inversely proportional to the square of the Brillouin shift [Eq. \ref{eq:gain}] for comparable overlap between the electrostriction force and the acoustic wave. For instance, a BSBS gain of $2.1 \times 10^2$ m$^{-1}$W$^{-1}$ with a Brillouin shift of 7.7~GHz has been reported in the same material using a waveguide geometry \cite{pant2011chip}. Even with a significantly lower overlap between the optical and acoustic modes in a microsphere as compared to a waveguide, the simulated FSBS gain was over three orders of magnitude higher than the reported BSBS gain.

The calculated gain was used in a theoretical model for the Brillouin threshold\cite{mirnaziry2017lasing}. The model showed excellent agreement with the experimental data [Fig.~\ref{fig:data125u}(c)]. The threshold for Brillouin lasing was predicted from the model at 910 $\pm$ 20 $\upmu$W, close to the 'knee' at 0.76 mW observed in Fig. \ref{fig:threshold_linear}. Beyond the Brillouin lasing threshold, cascaded Stokes lines were observed. Previous theoretical work has established that anti-Stokes beams will be generated along with Stokes beams in a waveguide that supports cascaded forward Brillouin scattering\cite{wolff2017cascaded}. The analysis also applies to resonators that are simultaneously resonant for the pump, Stokes, anti-Stokes and acoustic waves. We confirmed the existence of anti-Stokes beams using an optical heterodyne method (see the supplementary material). As the theory agrees well with the experimental results, we can reliably confirm that we are observing cascaded intramodal forward Brillouin scattering in our resonators.

Soliton states or other low-threshold nonlinear processes were not observed due to  normal dispersion both from the material \cite{bae2020chip} and the geometry~\cite{jin2017dispersion}. Due to the high nonlinearity of chalcogenide glass, symmetry breaking is known to occur at powers of the order of a mW \cite{zhu2019nonlinear}. The damage threshold for chalcogenide microresonators is also known to be in the same range \cite{zhu2020photo}. Therefore, it is likely that the power of the comb from a microsphere resonator cannot be increased dramatically without using an external amplifier. However, we note that there is room for improvement in the overlap integral between the electrostrictive force and the acoustic field in Eq. \ref{eq:gain}. It is possible that a higher overlap integral could be realized using a different resonator geometry that confines the standing acoustic wave more tightly (possibly a wedge or disc resonator \cite{kang2017high}), leading to Brillouin combs of higher optical power. It has also been reported that the Brillouin gain can be enhanced in sub-wavelength waveguides where the radiation pressure at the air-waveguide interface dominate the electrostrictive forces \cite{kittlaus2016large,yu2018giant}. Exploring the prospects of cascaded FSBS combs in novel resonator geometries could be a direction for future research.

\section{Conclusion}

In conclusion, we demonstrate cascaded forward intramodal Brillouin scattering within a microresonator platform for the first time. We used a theoretical model to estimate the threshold for cascaded FSBS on a resonator and verified that it agrees with the experimental results. The resonator was excited using a tunable laser at 1550 nm and beat notes between the pump and Stokes beams were observed at multiples of 34.5 MHz, corresponding to 25 orders of Stokes beams. We also established the existence of anti-Stokes beams generated through FSBS via a heterodyne measurement. We note the applications of cascaded forward Brillouin scattering in gas sensors and photonic radio frequency sources. This work could lay the foundation for future work into the SBS phenomena in near and mid-infrared using chalcogenide optics.

\section*{Funding}
Air Force Office of  Scientific Research (FA9550-15-1-0506), \\
Office of Naval Research (N00014-19-1-2251, N00014-19-1-2382)

\begin{acknowledgments}
The authors gratefully acknowledge Lange Simmons (University of Colorado, Boulder), Dr. Omkar Supekar (University of Colorado, Boulder), Dr. Kyuyoung Bae (Radanta Corporation, Boulder, CO USA) and Dr. Jiangang Zhu (Deepsight Technology, San Fransisco, CA USA). T. Shanavas thanks Prof. Ethan Neil (University of Colorado, Boulder) for offering the computational physics class (PHYS 5070) at the University of Colorado, Boulder.
\end{acknowledgments}

\section*{Disclosures}
The authors declare no conflicts of interest.

\section*{Data Availability Statement}
The data that support the findings of this study are available from the corresponding author upon reasonable request.

\bibliography{aipsamp}

\providecommand{\noopsort}[1]{}\providecommand{\singleletter}[1]{#1}%
\begin{thebibliography}{47}%
\makeatletter
\providecommand \@ifxundefined [1]{%
 \@ifx{#1\undefined}
}%
\providecommand \@ifnum [1]{%
 \ifnum #1\expandafter \@firstoftwo
 \else \expandafter \@secondoftwo
 \fi
}%
\providecommand \@ifx [1]{%
 \ifx #1\expandafter \@firstoftwo
 \else \expandafter \@secondoftwo
 \fi
}%
\providecommand \natexlab [1]{#1}%
\providecommand \enquote  [1]{``#1''}%
\providecommand \bibnamefont  [1]{#1}%
\providecommand \bibfnamefont [1]{#1}%
\providecommand \citenamefont [1]{#1}%
\providecommand \href@noop [0]{\@secondoftwo}%
\providecommand \href [0]{\begingroup \@sanitize@url \@href}%
\providecommand \@href[1]{\@@startlink{#1}\@@href}%
\providecommand \@@href[1]{\endgroup#1\@@endlink}%
\providecommand \@sanitize@url [0]{\catcode `\\12\catcode `\$12\catcode
  `\&12\catcode `\#12\catcode `\^12\catcode `\_12\catcode `\%12\relax}%
\providecommand \@@startlink[1]{}%
\providecommand \@@endlink[0]{}%
\providecommand \url  [0]{\begingroup\@sanitize@url \@url }%
\providecommand \@url [1]{\endgroup\@href {#1}{\urlprefix }}%
\providecommand \urlprefix  [0]{URL }%
\providecommand \Eprint [0]{\href }%
\providecommand \doibase [0]{http://dx.doi.org/}%
\providecommand \selectlanguage [0]{\@gobble}%
\providecommand \bibinfo  [0]{\@secondoftwo}%
\providecommand \bibfield  [0]{\@secondoftwo}%
\providecommand \translation [1]{[#1]}%
\providecommand \BibitemOpen [0]{}%
\providecommand \bibitemStop [0]{}%
\providecommand \bibitemNoStop [0]{.\EOS\space}%
\providecommand \EOS [0]{\spacefactor3000\relax}%
\providecommand \BibitemShut  [1]{\csname bibitem#1\endcsname}%
\let\auto@bib@innerbib\@empty
\bibitem [{\citenamefont {Petit}\ \emph {et~al.}(2009)\citenamefont {Petit},
  \citenamefont {Carlie}, \citenamefont {Chen}, \citenamefont {Gaylord},
  \citenamefont {Massera}, \citenamefont {Boudebs}, \citenamefont {Hu},
  \citenamefont {Agarwal}, \citenamefont {Kimerling},\ and\ \citenamefont
  {Richardson}}]{petit2009compositional}%
  \BibitemOpen
  \bibfield  {author} {\bibinfo {author} {\bibfnamefont {L.}~\bibnamefont
  {Petit}}, \bibinfo {author} {\bibfnamefont {N.}~\bibnamefont {Carlie}},
  \bibinfo {author} {\bibfnamefont {H.}~\bibnamefont {Chen}}, \bibinfo {author}
  {\bibfnamefont {S.}~\bibnamefont {Gaylord}}, \bibinfo {author} {\bibfnamefont
  {J.}~\bibnamefont {Massera}}, \bibinfo {author} {\bibfnamefont
  {G.}~\bibnamefont {Boudebs}}, \bibinfo {author} {\bibfnamefont
  {J.}~\bibnamefont {Hu}}, \bibinfo {author} {\bibfnamefont {A.}~\bibnamefont
  {Agarwal}}, \bibinfo {author} {\bibfnamefont {L.}~\bibnamefont {Kimerling}},
  \ and\ \bibinfo {author} {\bibfnamefont {K.}~\bibnamefont {Richardson}},\
  }\bibfield  {title} {\enquote {\bibinfo {title} {Compositional dependence of
  the nonlinear refractive index of new germanium-based chalcogenide
  glasses},}\ }\href@noop {} {\bibfield  {journal} {\bibinfo  {journal}
  {Journal of Solid State Chemistry}\ }\textbf {\bibinfo {volume} {182}},\
  \bibinfo {pages} {2756--2761} (\bibinfo {year} {2009})}\BibitemShut {NoStop}%
\bibitem [{\citenamefont {Eggleton}, \citenamefont {Luther-Davies},\ and\
  \citenamefont {Richardson}(2011)}]{eggleton2011chalcogenide}%
  \BibitemOpen
  \bibfield  {author} {\bibinfo {author} {\bibfnamefont {B.~J.}\ \bibnamefont
  {Eggleton}}, \bibinfo {author} {\bibfnamefont {B.}~\bibnamefont
  {Luther-Davies}}, \ and\ \bibinfo {author} {\bibfnamefont {K.}~\bibnamefont
  {Richardson}},\ }\bibfield  {title} {\enquote {\bibinfo {title} {Chalcogenide
  photonics},}\ }\href@noop {} {\bibfield  {journal} {\bibinfo  {journal}
  {Nature photonics}\ }\textbf {\bibinfo {volume} {5}},\ \bibinfo {pages}
  {141--148} (\bibinfo {year} {2011})}\BibitemShut {NoStop}%
\bibitem [{\citenamefont {Zakery}\ and\ \citenamefont
  {Elliott}(2003)}]{zakery2003optical}%
  \BibitemOpen
  \bibfield  {author} {\bibinfo {author} {\bibfnamefont {A.}~\bibnamefont
  {Zakery}}\ and\ \bibinfo {author} {\bibfnamefont {S.}~\bibnamefont
  {Elliott}},\ }\bibfield  {title} {\enquote {\bibinfo {title} {Optical
  properties and applications of chalcogenide glasses: a review},}\ }\href@noop
  {} {\bibfield  {journal} {\bibinfo  {journal} {Journal of Non-Crystalline
  Solids}\ }\textbf {\bibinfo {volume} {330}},\ \bibinfo {pages} {1--12}
  (\bibinfo {year} {2003})}\BibitemShut {NoStop}%
\bibitem [{\citenamefont {Scheer}\ and\ \citenamefont
  {Schock}(2011)}]{scheer2011chalcogenide}%
  \BibitemOpen
  \bibfield  {author} {\bibinfo {author} {\bibfnamefont {R.}~\bibnamefont
  {Scheer}}\ and\ \bibinfo {author} {\bibfnamefont {H.-W.}\ \bibnamefont
  {Schock}},\ }\href@noop {} {\emph {\bibinfo {title} {Chalcogenide
  photovoltaics: physics, technologies, and thin film devices}}}\ (\bibinfo
  {publisher} {John Wiley \& Sons},\ \bibinfo {year} {2011})\BibitemShut
  {NoStop}%
\bibitem [{\citenamefont {Adam}\ and\ \citenamefont
  {Zhang}(2014)}]{adam2014chalcogenide}%
  \BibitemOpen
  \bibfield  {author} {\bibinfo {author} {\bibfnamefont {J.-L.}\ \bibnamefont
  {Adam}}\ and\ \bibinfo {author} {\bibfnamefont {X.}~\bibnamefont {Zhang}},\
  }\href@noop {} {\emph {\bibinfo {title} {Chalcogenide glasses: preparation,
  properties and applications}}}\ (\bibinfo  {publisher} {Woodhead
  publishing},\ \bibinfo {year} {2014})\BibitemShut {NoStop}%
\bibitem [{\citenamefont {Lyubin}\ \emph {et~al.}(1997)\citenamefont {Lyubin},
  \citenamefont {Klebanov}, \citenamefont {Mitkova},\ and\ \citenamefont
  {Petkova}}]{lyubin1997polarization}%
  \BibitemOpen
  \bibfield  {author} {\bibinfo {author} {\bibfnamefont {V.}~\bibnamefont
  {Lyubin}}, \bibinfo {author} {\bibfnamefont {M.}~\bibnamefont {Klebanov}},
  \bibinfo {author} {\bibfnamefont {M.}~\bibnamefont {Mitkova}}, \ and\
  \bibinfo {author} {\bibfnamefont {T.}~\bibnamefont {Petkova}},\ }\bibfield
  {title} {\enquote {\bibinfo {title} {Polarization-dependent, laser-induced
  anisotropic photocrystallization of some amorphous chalcogenide films},}\
  }\href@noop {} {\bibfield  {journal} {\bibinfo  {journal} {Applied physics
  letters}\ }\textbf {\bibinfo {volume} {71}},\ \bibinfo {pages} {2118--2120}
  (\bibinfo {year} {1997})}\BibitemShut {NoStop}%
\bibitem [{\citenamefont {Fritzsche}(1993)}]{fritzsche1993origin}%
  \BibitemOpen
  \bibfield  {author} {\bibinfo {author} {\bibfnamefont {H.}~\bibnamefont
  {Fritzsche}},\ }\bibfield  {title} {\enquote {\bibinfo {title} {The origin of
  reversible and irreversible photostructural changes in chalcogenide
  glasses},}\ }\href@noop {} {\bibfield  {journal} {\bibinfo  {journal}
  {Philosophical Magazine B}\ }\textbf {\bibinfo {volume} {68}},\ \bibinfo
  {pages} {561--572} (\bibinfo {year} {1993})}\BibitemShut {NoStop}%
\bibitem [{\citenamefont {Owen}, \citenamefont {Firth},\ and\ \citenamefont
  {Ewen}(1985)}]{owen1985photo}%
  \BibitemOpen
  \bibfield  {author} {\bibinfo {author} {\bibfnamefont {A.}~\bibnamefont
  {Owen}}, \bibinfo {author} {\bibfnamefont {A.}~\bibnamefont {Firth}}, \ and\
  \bibinfo {author} {\bibfnamefont {P.}~\bibnamefont {Ewen}},\ }\bibfield
  {title} {\enquote {\bibinfo {title} {Photo-induced structural and
  physico-chemical changes in amorphous chalcogenide semiconductors},}\
  }\href@noop {} {\bibfield  {journal} {\bibinfo  {journal} {Philosophical
  Magazine B}\ }\textbf {\bibinfo {volume} {52}},\ \bibinfo {pages} {347--362}
  (\bibinfo {year} {1985})}\BibitemShut {NoStop}%
\bibitem [{\citenamefont {Chopra}\ \emph {et~al.}(1981)\citenamefont {Chopra},
  \citenamefont {Harshvardhan}, \citenamefont {Rajagopalan},\ and\
  \citenamefont {Malhotra}}]{chopra1981origin}%
  \BibitemOpen
  \bibfield  {author} {\bibinfo {author} {\bibfnamefont {K.}~\bibnamefont
  {Chopra}}, \bibinfo {author} {\bibfnamefont {K.~S.}\ \bibnamefont
  {Harshvardhan}}, \bibinfo {author} {\bibfnamefont {S.}~\bibnamefont
  {Rajagopalan}}, \ and\ \bibinfo {author} {\bibfnamefont {L.}~\bibnamefont
  {Malhotra}},\ }\bibfield  {title} {\enquote {\bibinfo {title} {On the origin
  of photocontraction effect in amorphous chalcogenide films},}\ }\href@noop {}
  {\bibfield  {journal} {\bibinfo  {journal} {Solid State Communications}\
  }\textbf {\bibinfo {volume} {40}},\ \bibinfo {pages} {387--390} (\bibinfo
  {year} {1981})}\BibitemShut {NoStop}%
\bibitem [{\citenamefont {Seddon}(1995)}]{seddon1995chalcogenide}%
  \BibitemOpen
  \bibfield  {author} {\bibinfo {author} {\bibfnamefont {A.}~\bibnamefont
  {Seddon}},\ }\bibfield  {title} {\enquote {\bibinfo {title} {Chalcogenide
  glasses: a review of their preparation, properties and applications},}\
  }\href@noop {} {\bibfield  {journal} {\bibinfo  {journal} {Journal of
  Non-Crystalline Solids}\ }\textbf {\bibinfo {volume} {184}},\ \bibinfo
  {pages} {44--50} (\bibinfo {year} {1995})}\BibitemShut {NoStop}%
\bibitem [{\citenamefont {Bae}\ \emph {et~al.}(2020)\citenamefont {Bae},
  \citenamefont {Horning}, \citenamefont {Pampel}, \citenamefont {Zohrabi},
  \citenamefont {Grayson}, \citenamefont {Gopinath},\ and\ \citenamefont
  {Park}}]{bae2020chip}%
  \BibitemOpen
  \bibfield  {author} {\bibinfo {author} {\bibfnamefont {K.}~\bibnamefont
  {Bae}}, \bibinfo {author} {\bibfnamefont {T.~M.}\ \bibnamefont {Horning}},
  \bibinfo {author} {\bibfnamefont {S.}~\bibnamefont {Pampel}}, \bibinfo
  {author} {\bibfnamefont {M.}~\bibnamefont {Zohrabi}}, \bibinfo {author}
  {\bibfnamefont {M.~B.}\ \bibnamefont {Grayson}}, \bibinfo {author}
  {\bibfnamefont {J.~T.}\ \bibnamefont {Gopinath}}, \ and\ \bibinfo {author}
  {\bibfnamefont {W.}~\bibnamefont {Park}},\ }\bibfield  {title} {\enquote
  {\bibinfo {title} {On-chip high-quality ge 23 sb 7 s 70 round-wedge
  resonators for broadband dispersion engineering},}\ }in\ \href@noop {} {\emph
  {\bibinfo {booktitle} {2020 Conference on Lasers and Electro-Optics
  (CLEO)}}}\ (\bibinfo {organization} {IEEE},\ \bibinfo {year} {2020})\ pp.\
  \bibinfo {pages} {1--2}\BibitemShut {NoStop}%
\bibitem [{\citenamefont {Hu}\ \emph {et~al.}(2010)\citenamefont {Hu},
  \citenamefont {Feng}, \citenamefont {Carlie}, \citenamefont {Petit},
  \citenamefont {Agarwal}, \citenamefont {Richardson},\ and\ \citenamefont
  {Kimerling}}]{hu2010optical}%
  \BibitemOpen
  \bibfield  {author} {\bibinfo {author} {\bibfnamefont {J.}~\bibnamefont
  {Hu}}, \bibinfo {author} {\bibfnamefont {N.-N.}\ \bibnamefont {Feng}},
  \bibinfo {author} {\bibfnamefont {N.}~\bibnamefont {Carlie}}, \bibinfo
  {author} {\bibfnamefont {L.}~\bibnamefont {Petit}}, \bibinfo {author}
  {\bibfnamefont {A.}~\bibnamefont {Agarwal}}, \bibinfo {author} {\bibfnamefont
  {K.}~\bibnamefont {Richardson}}, \ and\ \bibinfo {author} {\bibfnamefont
  {L.}~\bibnamefont {Kimerling}},\ }\bibfield  {title} {\enquote {\bibinfo
  {title} {Optical loss reduction in high-index-contrast chalcogenide glass
  waveguides via thermal reflow},}\ }\href@noop {} {\bibfield  {journal}
  {\bibinfo  {journal} {Optics Express}\ }\textbf {\bibinfo {volume} {18}},\
  \bibinfo {pages} {1469--1478} (\bibinfo {year} {2010})}\BibitemShut {NoStop}%
\bibitem [{\citenamefont {Zhao}\ \emph {et~al.}(2019)\citenamefont {Zhao},
  \citenamefont {Li}, \citenamefont {Guo}, \citenamefont {Zhang}, \citenamefont
  {Xu},\ and\ \citenamefont {Zhang}}]{zhao2019exploration}%
  \BibitemOpen
  \bibfield  {author} {\bibinfo {author} {\bibfnamefont {Y.}~\bibnamefont
  {Zhao}}, \bibinfo {author} {\bibfnamefont {C.}~\bibnamefont {Li}}, \bibinfo
  {author} {\bibfnamefont {P.}~\bibnamefont {Guo}}, \bibinfo {author}
  {\bibfnamefont {W.}~\bibnamefont {Zhang}}, \bibinfo {author} {\bibfnamefont
  {P.}~\bibnamefont {Xu}}, \ and\ \bibinfo {author} {\bibfnamefont
  {P.}~\bibnamefont {Zhang}},\ }\bibfield  {title} {\enquote {\bibinfo {title}
  {Exploration of lift-off ge--as--se chalcogenide waveguides with thermal
  reflow process},}\ }\href@noop {} {\bibfield  {journal} {\bibinfo  {journal}
  {Optical Materials}\ }\textbf {\bibinfo {volume} {92}},\ \bibinfo {pages}
  {206--211} (\bibinfo {year} {2019})}\BibitemShut {NoStop}%
\bibitem [{\citenamefont {Loh}\ \emph {et~al.}(2019)\citenamefont {Loh},
  \citenamefont {Yegnanarayanan}, \citenamefont {O’Donnell},\ and\
  \citenamefont {Juodawlkis}}]{loh2019ultra}%
  \BibitemOpen
  \bibfield  {author} {\bibinfo {author} {\bibfnamefont {W.}~\bibnamefont
  {Loh}}, \bibinfo {author} {\bibfnamefont {S.}~\bibnamefont {Yegnanarayanan}},
  \bibinfo {author} {\bibfnamefont {F.}~\bibnamefont {O’Donnell}}, \ and\
  \bibinfo {author} {\bibfnamefont {P.~W.}\ \bibnamefont {Juodawlkis}},\
  }\bibfield  {title} {\enquote {\bibinfo {title} {Ultra-narrow linewidth
  brillouin laser with nanokelvin temperature self-referencing},}\ }\href@noop
  {} {\bibfield  {journal} {\bibinfo  {journal} {Optica}\ }\textbf {\bibinfo
  {volume} {6}},\ \bibinfo {pages} {152--159} (\bibinfo {year}
  {2019})}\BibitemShut {NoStop}%
\bibitem [{\citenamefont {Loh}\ \emph {et~al.}(2015)\citenamefont {Loh},
  \citenamefont {Green}, \citenamefont {Baynes}, \citenamefont {Cole},
  \citenamefont {Quinlan}, \citenamefont {Lee}, \citenamefont {Vahala},
  \citenamefont {Papp},\ and\ \citenamefont {Diddams}}]{loh2015dual}%
  \BibitemOpen
  \bibfield  {author} {\bibinfo {author} {\bibfnamefont {W.}~\bibnamefont
  {Loh}}, \bibinfo {author} {\bibfnamefont {A.~A.}\ \bibnamefont {Green}},
  \bibinfo {author} {\bibfnamefont {F.~N.}\ \bibnamefont {Baynes}}, \bibinfo
  {author} {\bibfnamefont {D.~C.}\ \bibnamefont {Cole}}, \bibinfo {author}
  {\bibfnamefont {F.~J.}\ \bibnamefont {Quinlan}}, \bibinfo {author}
  {\bibfnamefont {H.}~\bibnamefont {Lee}}, \bibinfo {author} {\bibfnamefont
  {K.~J.}\ \bibnamefont {Vahala}}, \bibinfo {author} {\bibfnamefont {S.~B.}\
  \bibnamefont {Papp}}, \ and\ \bibinfo {author} {\bibfnamefont {S.~A.}\
  \bibnamefont {Diddams}},\ }\bibfield  {title} {\enquote {\bibinfo {title}
  {Dual-microcavity narrow-linewidth brillouin laser},}\ }\href@noop {}
  {\bibfield  {journal} {\bibinfo  {journal} {Optica}\ }\textbf {\bibinfo
  {volume} {2}},\ \bibinfo {pages} {225--232} (\bibinfo {year}
  {2015})}\BibitemShut {NoStop}%
\bibitem [{\citenamefont {Zadok}, \citenamefont {Eyal},\ and\ \citenamefont
  {Tur}(2011)}]{zadok2011stimulated}%
  \BibitemOpen
  \bibfield  {author} {\bibinfo {author} {\bibfnamefont {A.}~\bibnamefont
  {Zadok}}, \bibinfo {author} {\bibfnamefont {A.}~\bibnamefont {Eyal}}, \ and\
  \bibinfo {author} {\bibfnamefont {M.}~\bibnamefont {Tur}},\ }\bibfield
  {title} {\enquote {\bibinfo {title} {Stimulated brillouin scattering slow
  light in optical fibers},}\ }\href@noop {} {\bibfield  {journal} {\bibinfo
  {journal} {Applied Optics}\ }\textbf {\bibinfo {volume} {50}},\ \bibinfo
  {pages} {E38--E49} (\bibinfo {year} {2011})}\BibitemShut {NoStop}%
\bibitem [{\citenamefont {Bahl}\ \emph {et~al.}(2012)\citenamefont {Bahl},
  \citenamefont {Tomes}, \citenamefont {Marquardt},\ and\ \citenamefont
  {Carmon}}]{bahl2012observation}%
  \BibitemOpen
  \bibfield  {author} {\bibinfo {author} {\bibfnamefont {G.}~\bibnamefont
  {Bahl}}, \bibinfo {author} {\bibfnamefont {M.}~\bibnamefont {Tomes}},
  \bibinfo {author} {\bibfnamefont {F.}~\bibnamefont {Marquardt}}, \ and\
  \bibinfo {author} {\bibfnamefont {T.}~\bibnamefont {Carmon}},\ }\bibfield
  {title} {\enquote {\bibinfo {title} {Observation of spontaneous brillouin
  cooling},}\ }\href@noop {} {\bibfield  {journal} {\bibinfo  {journal} {Nature
  Physics}\ }\textbf {\bibinfo {volume} {8}},\ \bibinfo {pages} {203--207}
  (\bibinfo {year} {2012})}\BibitemShut {NoStop}%
\bibitem [{\citenamefont {Ma}\ \emph {et~al.}(2020)\citenamefont {Ma},
  \citenamefont {Wen}, \citenamefont {Ding}, \citenamefont {Li}, \citenamefont
  {Hu}, \citenamefont {Jiang}, \citenamefont {Jiang},\ and\ \citenamefont
  {Xiao}}]{ma2020chip}%
  \BibitemOpen
  \bibfield  {author} {\bibinfo {author} {\bibfnamefont {J.}~\bibnamefont
  {Ma}}, \bibinfo {author} {\bibfnamefont {J.}~\bibnamefont {Wen}}, \bibinfo
  {author} {\bibfnamefont {S.}~\bibnamefont {Ding}}, \bibinfo {author}
  {\bibfnamefont {S.}~\bibnamefont {Li}}, \bibinfo {author} {\bibfnamefont
  {Y.}~\bibnamefont {Hu}}, \bibinfo {author} {\bibfnamefont {X.}~\bibnamefont
  {Jiang}}, \bibinfo {author} {\bibfnamefont {L.}~\bibnamefont {Jiang}}, \ and\
  \bibinfo {author} {\bibfnamefont {M.}~\bibnamefont {Xiao}},\ }\bibfield
  {title} {\enquote {\bibinfo {title} {Chip-based optical isolator and
  nonreciprocal parity-time symmetry induced by stimulated brillouin
  scattering},}\ }\href@noop {} {\bibfield  {journal} {\bibinfo  {journal}
  {Laser \& Photonics Reviews}\ }\textbf {\bibinfo {volume} {14}},\ \bibinfo
  {pages} {1900278} (\bibinfo {year} {2020})}\BibitemShut {NoStop}%
\bibitem [{\citenamefont {Horiguchi}\ \emph {et~al.}(1995)\citenamefont
  {Horiguchi}, \citenamefont {Shimizu}, \citenamefont {Kurashima},
  \citenamefont {Tateda},\ and\ \citenamefont
  {Koyamada}}]{horiguchi1995development}%
  \BibitemOpen
  \bibfield  {author} {\bibinfo {author} {\bibfnamefont {T.}~\bibnamefont
  {Horiguchi}}, \bibinfo {author} {\bibfnamefont {K.}~\bibnamefont {Shimizu}},
  \bibinfo {author} {\bibfnamefont {T.}~\bibnamefont {Kurashima}}, \bibinfo
  {author} {\bibfnamefont {M.}~\bibnamefont {Tateda}}, \ and\ \bibinfo {author}
  {\bibfnamefont {Y.}~\bibnamefont {Koyamada}},\ }\bibfield  {title} {\enquote
  {\bibinfo {title} {Development of a distributed sensing technique using
  brillouin scattering},}\ }\href@noop {} {\bibfield  {journal} {\bibinfo
  {journal} {Journal of lightwave technology}\ }\textbf {\bibinfo {volume}
  {13}},\ \bibinfo {pages} {1296--1302} (\bibinfo {year} {1995})}\BibitemShut
  {NoStop}%
\bibitem [{\citenamefont {Qiu}\ \emph {et~al.}(2013)\citenamefont {Qiu},
  \citenamefont {Rakich}, \citenamefont {Shin}, \citenamefont {Dong},
  \citenamefont {Solja{\v{c}}i{\'c}},\ and\ \citenamefont
  {Wang}}]{qiu2013stimulated}%
  \BibitemOpen
  \bibfield  {author} {\bibinfo {author} {\bibfnamefont {W.}~\bibnamefont
  {Qiu}}, \bibinfo {author} {\bibfnamefont {P.~T.}\ \bibnamefont {Rakich}},
  \bibinfo {author} {\bibfnamefont {H.}~\bibnamefont {Shin}}, \bibinfo {author}
  {\bibfnamefont {H.}~\bibnamefont {Dong}}, \bibinfo {author} {\bibfnamefont
  {M.}~\bibnamefont {Solja{\v{c}}i{\'c}}}, \ and\ \bibinfo {author}
  {\bibfnamefont {Z.}~\bibnamefont {Wang}},\ }\bibfield  {title} {\enquote
  {\bibinfo {title} {Stimulated brillouin scattering in nanoscale silicon
  step-index waveguides: a general framework of selection rules and calculating
  sbs gain},}\ }\href@noop {} {\bibfield  {journal} {\bibinfo  {journal}
  {Optics express}\ }\textbf {\bibinfo {volume} {21}},\ \bibinfo {pages}
  {31402--31419} (\bibinfo {year} {2013})}\BibitemShut {NoStop}%
\bibitem [{\citenamefont {Hill}, \citenamefont {Kawasaki},\ and\ \citenamefont
  {Johnson}(1976)}]{hill1976cw}%
  \BibitemOpen
  \bibfield  {author} {\bibinfo {author} {\bibfnamefont {K.}~\bibnamefont
  {Hill}}, \bibinfo {author} {\bibfnamefont {B.}~\bibnamefont {Kawasaki}}, \
  and\ \bibinfo {author} {\bibfnamefont {D.}~\bibnamefont {Johnson}},\
  }\bibfield  {title} {\enquote {\bibinfo {title} {Cw brillouin laser},}\
  }\href@noop {} {\bibfield  {journal} {\bibinfo  {journal} {Applied Physics
  Letters}\ }\textbf {\bibinfo {volume} {28}},\ \bibinfo {pages} {608--609}
  (\bibinfo {year} {1976})}\BibitemShut {NoStop}%
\bibitem [{\citenamefont {Wang}\ \emph {et~al.}(2021)\citenamefont {Wang},
  \citenamefont {Yu}, \citenamefont {Li}, \citenamefont {Bai}, \citenamefont
  {Wang}, \citenamefont {Li}, \citenamefont {Song}, \citenamefont {Wang},
  \citenamefont {Li}, \citenamefont {Wang} \emph
  {et~al.}}]{wang2021tailorable}%
  \BibitemOpen
  \bibfield  {author} {\bibinfo {author} {\bibfnamefont {W.}~\bibnamefont
  {Wang}}, \bibinfo {author} {\bibfnamefont {Y.}~\bibnamefont {Yu}}, \bibinfo
  {author} {\bibfnamefont {Y.}~\bibnamefont {Li}}, \bibinfo {author}
  {\bibfnamefont {Z.}~\bibnamefont {Bai}}, \bibinfo {author} {\bibfnamefont
  {G.}~\bibnamefont {Wang}}, \bibinfo {author} {\bibfnamefont {K.}~\bibnamefont
  {Li}}, \bibinfo {author} {\bibfnamefont {C.}~\bibnamefont {Song}}, \bibinfo
  {author} {\bibfnamefont {Z.}~\bibnamefont {Wang}}, \bibinfo {author}
  {\bibfnamefont {S.}~\bibnamefont {Li}}, \bibinfo {author} {\bibfnamefont
  {Y.}~\bibnamefont {Wang}},  \emph {et~al.},\ }\bibfield  {title} {\enquote
  {\bibinfo {title} {Tailorable brillouin light scattering in a lithium niobate
  waveguide},}\ }\href@noop {} {\bibfield  {journal} {\bibinfo  {journal}
  {Applied Sciences}\ }\textbf {\bibinfo {volume} {11}},\ \bibinfo {pages}
  {8390} (\bibinfo {year} {2021})}\BibitemShut {NoStop}%
\bibitem [{\citenamefont {Mirnaziry}\ \emph {et~al.}(2017)\citenamefont
  {Mirnaziry}, \citenamefont {Wolff}, \citenamefont {Steel}, \citenamefont
  {Morrison}, \citenamefont {Eggleton},\ and\ \citenamefont
  {Poulton}}]{mirnaziry2017lasing}%
  \BibitemOpen
  \bibfield  {author} {\bibinfo {author} {\bibfnamefont {S.~R.}\ \bibnamefont
  {Mirnaziry}}, \bibinfo {author} {\bibfnamefont {C.}~\bibnamefont {Wolff}},
  \bibinfo {author} {\bibfnamefont {M.}~\bibnamefont {Steel}}, \bibinfo
  {author} {\bibfnamefont {B.}~\bibnamefont {Morrison}}, \bibinfo {author}
  {\bibfnamefont {B.~J.}\ \bibnamefont {Eggleton}}, \ and\ \bibinfo {author}
  {\bibfnamefont {C.~G.}\ \bibnamefont {Poulton}},\ }\bibfield  {title}
  {\enquote {\bibinfo {title} {Lasing in ring resonators by stimulated
  brillouin scattering in the presence of nonlinear loss},}\ }\href@noop {}
  {\bibfield  {journal} {\bibinfo  {journal} {Optics Express}\ }\textbf
  {\bibinfo {volume} {25}},\ \bibinfo {pages} {23619--23633} (\bibinfo {year}
  {2017})}\BibitemShut {NoStop}%
\bibitem [{\citenamefont {Lin}\ and\ \citenamefont
  {Chembo}(2016)}]{lin2016opto}%
  \BibitemOpen
  \bibfield  {author} {\bibinfo {author} {\bibfnamefont {G.}~\bibnamefont
  {Lin}}\ and\ \bibinfo {author} {\bibfnamefont {Y.~K.}\ \bibnamefont
  {Chembo}},\ }\bibfield  {title} {\enquote {\bibinfo {title} {Opto-acoustic
  phenomena in whispering gallery mode resonators},}\ }\href@noop {} {\bibfield
   {journal} {\bibinfo  {journal} {International Journal of Optomechatronics}\
  }\textbf {\bibinfo {volume} {10}},\ \bibinfo {pages} {32--39} (\bibinfo
  {year} {2016})}\BibitemShut {NoStop}%
\bibitem [{\citenamefont {Del'Haye}, \citenamefont {Diddams},\ and\
  \citenamefont {Papp}(2013)}]{del2013laser}%
  \BibitemOpen
  \bibfield  {author} {\bibinfo {author} {\bibfnamefont {P.}~\bibnamefont
  {Del'Haye}}, \bibinfo {author} {\bibfnamefont {S.~A.}\ \bibnamefont
  {Diddams}}, \ and\ \bibinfo {author} {\bibfnamefont {S.~B.}\ \bibnamefont
  {Papp}},\ }\bibfield  {title} {\enquote {\bibinfo {title} {Laser-machined
  ultra-high-q microrod resonators for nonlinear optics},}\ }\href@noop {}
  {\bibfield  {journal} {\bibinfo  {journal} {Applied Physics Letters}\
  }\textbf {\bibinfo {volume} {102}},\ \bibinfo {pages} {221119} (\bibinfo
  {year} {2013})}\BibitemShut {NoStop}%
\bibitem [{\citenamefont {Grudinin}, \citenamefont {Matsko},\ and\
  \citenamefont {Maleki}(2009)}]{grudinin2009brillouin}%
  \BibitemOpen
  \bibfield  {author} {\bibinfo {author} {\bibfnamefont {I.~S.}\ \bibnamefont
  {Grudinin}}, \bibinfo {author} {\bibfnamefont {A.~B.}\ \bibnamefont
  {Matsko}}, \ and\ \bibinfo {author} {\bibfnamefont {L.}~\bibnamefont
  {Maleki}},\ }\bibfield  {title} {\enquote {\bibinfo {title} {Brillouin lasing
  with a caf 2 whispering gallery mode resonator},}\ }\href@noop {} {\bibfield
  {journal} {\bibinfo  {journal} {Physical review letters}\ }\textbf {\bibinfo
  {volume} {102}},\ \bibinfo {pages} {043902} (\bibinfo {year}
  {2009})}\BibitemShut {NoStop}%
\bibitem [{\citenamefont {Yu}\ \emph {et~al.}(2022)\citenamefont {Yu},
  \citenamefont {Shen}, \citenamefont {Yang}, \citenamefont {Qi}, \citenamefont
  {Jiang}, \citenamefont {Brambilla}, \citenamefont {Dong},\ and\ \citenamefont
  {Wang}}]{yu2022investigation}%
  \BibitemOpen
  \bibfield  {author} {\bibinfo {author} {\bibfnamefont {J.}~\bibnamefont
  {Yu}}, \bibinfo {author} {\bibfnamefont {Z.}~\bibnamefont {Shen}}, \bibinfo
  {author} {\bibfnamefont {Z.}~\bibnamefont {Yang}}, \bibinfo {author}
  {\bibfnamefont {S.}~\bibnamefont {Qi}}, \bibinfo {author} {\bibfnamefont
  {Y.}~\bibnamefont {Jiang}}, \bibinfo {author} {\bibfnamefont
  {G.}~\bibnamefont {Brambilla}}, \bibinfo {author} {\bibfnamefont {C.-H.}\
  \bibnamefont {Dong}}, \ and\ \bibinfo {author} {\bibfnamefont
  {P.}~\bibnamefont {Wang}},\ }\bibfield  {title} {\enquote {\bibinfo {title}
  {The investigation of forward and backward brillouin scattering in high-q
  chalcogenide microspheres},}\ }\href@noop {} {\bibfield  {journal} {\bibinfo
  {journal} {IEEE Photonics Journal}\ }\textbf {\bibinfo {volume} {14}},\
  \bibinfo {pages} {1--5} (\bibinfo {year} {2022})}\BibitemShut {NoStop}%
\bibitem [{\citenamefont {Bahl}\ \emph {et~al.}(2011)\citenamefont {Bahl},
  \citenamefont {Zehnpfennig}, \citenamefont {Tomes},\ and\ \citenamefont
  {Carmon}}]{bahl2011stimulated}%
  \BibitemOpen
  \bibfield  {author} {\bibinfo {author} {\bibfnamefont {G.}~\bibnamefont
  {Bahl}}, \bibinfo {author} {\bibfnamefont {J.}~\bibnamefont {Zehnpfennig}},
  \bibinfo {author} {\bibfnamefont {M.}~\bibnamefont {Tomes}}, \ and\ \bibinfo
  {author} {\bibfnamefont {T.}~\bibnamefont {Carmon}},\ }\bibfield  {title}
  {\enquote {\bibinfo {title} {Stimulated optomechanical excitation of surface
  acoustic waves in a microdevice},}\ }\href@noop {} {\bibfield  {journal}
  {\bibinfo  {journal} {Nature communications}\ }\textbf {\bibinfo {volume}
  {2}},\ \bibinfo {pages} {1--6} (\bibinfo {year} {2011})}\BibitemShut
  {NoStop}%
\bibitem [{\citenamefont {Guo}\ \emph {et~al.}(2015)\citenamefont {Guo},
  \citenamefont {Che}, \citenamefont {Cai}, \citenamefont {Liu}, \citenamefont
  {Gu}, \citenamefont {Chu}, \citenamefont {Zhang}, \citenamefont {Fu},
  \citenamefont {Luo},\ and\ \citenamefont {Xu}}]{guo2015ultralow}%
  \BibitemOpen
  \bibfield  {author} {\bibinfo {author} {\bibfnamefont {C.}~\bibnamefont
  {Guo}}, \bibinfo {author} {\bibfnamefont {K.}~\bibnamefont {Che}}, \bibinfo
  {author} {\bibfnamefont {Z.}~\bibnamefont {Cai}}, \bibinfo {author}
  {\bibfnamefont {S.}~\bibnamefont {Liu}}, \bibinfo {author} {\bibfnamefont
  {G.}~\bibnamefont {Gu}}, \bibinfo {author} {\bibfnamefont {C.}~\bibnamefont
  {Chu}}, \bibinfo {author} {\bibfnamefont {P.}~\bibnamefont {Zhang}}, \bibinfo
  {author} {\bibfnamefont {H.}~\bibnamefont {Fu}}, \bibinfo {author}
  {\bibfnamefont {Z.}~\bibnamefont {Luo}}, \ and\ \bibinfo {author}
  {\bibfnamefont {H.}~\bibnamefont {Xu}},\ }\bibfield  {title} {\enquote
  {\bibinfo {title} {Ultralow-threshold cascaded brillouin microlaser for
  tunable microwave generation},}\ }\href@noop {} {\bibfield  {journal}
  {\bibinfo  {journal} {Optics Letters}\ }\textbf {\bibinfo {volume} {40}},\
  \bibinfo {pages} {4971--4974} (\bibinfo {year} {2015})}\BibitemShut {NoStop}%
\bibitem [{\citenamefont {Yao}\ \emph {et~al.}(2017)\citenamefont {Yao},
  \citenamefont {Yu}, \citenamefont {Wu}, \citenamefont {Huang}, \citenamefont
  {Wu}, \citenamefont {Gong}, \citenamefont {Chen}, \citenamefont {Li},
  \citenamefont {Wong}, \citenamefont {Fan} \emph {et~al.}}]{yao2017graphene}%
  \BibitemOpen
  \bibfield  {author} {\bibinfo {author} {\bibfnamefont {B.}~\bibnamefont
  {Yao}}, \bibinfo {author} {\bibfnamefont {C.}~\bibnamefont {Yu}}, \bibinfo
  {author} {\bibfnamefont {Y.}~\bibnamefont {Wu}}, \bibinfo {author}
  {\bibfnamefont {S.-W.}\ \bibnamefont {Huang}}, \bibinfo {author}
  {\bibfnamefont {H.}~\bibnamefont {Wu}}, \bibinfo {author} {\bibfnamefont
  {Y.}~\bibnamefont {Gong}}, \bibinfo {author} {\bibfnamefont {Y.}~\bibnamefont
  {Chen}}, \bibinfo {author} {\bibfnamefont {Y.}~\bibnamefont {Li}}, \bibinfo
  {author} {\bibfnamefont {C.~W.}\ \bibnamefont {Wong}}, \bibinfo {author}
  {\bibfnamefont {X.}~\bibnamefont {Fan}},  \emph {et~al.},\ }\bibfield
  {title} {\enquote {\bibinfo {title} {Graphene-enhanced brillouin
  optomechanical microresonator for ultrasensitive gas detection},}\
  }\href@noop {} {\bibfield  {journal} {\bibinfo  {journal} {Nano letters}\
  }\textbf {\bibinfo {volume} {17}},\ \bibinfo {pages} {4996--5002} (\bibinfo
  {year} {2017})}\BibitemShut {NoStop}%
\bibitem [{\citenamefont {Fortier}\ \emph {et~al.}(2011)\citenamefont
  {Fortier}, \citenamefont {Kirchner}, \citenamefont {Quinlan}, \citenamefont
  {Taylor}, \citenamefont {Bergquist}, \citenamefont {Rosenband}, \citenamefont
  {Lemke}, \citenamefont {Ludlow}, \citenamefont {Jiang}, \citenamefont {Oates}
  \emph {et~al.}}]{fortier2011generation}%
  \BibitemOpen
  \bibfield  {author} {\bibinfo {author} {\bibfnamefont {T.~M.}\ \bibnamefont
  {Fortier}}, \bibinfo {author} {\bibfnamefont {M.~S.}\ \bibnamefont
  {Kirchner}}, \bibinfo {author} {\bibfnamefont {F.}~\bibnamefont {Quinlan}},
  \bibinfo {author} {\bibfnamefont {J.}~\bibnamefont {Taylor}}, \bibinfo
  {author} {\bibfnamefont {J.}~\bibnamefont {Bergquist}}, \bibinfo {author}
  {\bibfnamefont {T.}~\bibnamefont {Rosenband}}, \bibinfo {author}
  {\bibfnamefont {N.}~\bibnamefont {Lemke}}, \bibinfo {author} {\bibfnamefont
  {A.}~\bibnamefont {Ludlow}}, \bibinfo {author} {\bibfnamefont
  {Y.}~\bibnamefont {Jiang}}, \bibinfo {author} {\bibfnamefont
  {C.}~\bibnamefont {Oates}},  \emph {et~al.},\ }\bibfield  {title} {\enquote
  {\bibinfo {title} {Generation of ultrastable microwaves via optical frequency
  division},}\ }\href@noop {} {\bibfield  {journal} {\bibinfo  {journal}
  {Nature Photonics}\ }\textbf {\bibinfo {volume} {5}},\ \bibinfo {pages}
  {425--429} (\bibinfo {year} {2011})}\BibitemShut {NoStop}%
\bibitem [{\citenamefont {Gundavarapu}\ \emph {et~al.}(2019)\citenamefont
  {Gundavarapu}, \citenamefont {Brodnik}, \citenamefont {Puckett},
  \citenamefont {Huffman}, \citenamefont {Bose}, \citenamefont {Behunin},
  \citenamefont {Wu}, \citenamefont {Qiu}, \citenamefont {Pinho}, \citenamefont
  {Chauhan} \emph {et~al.}}]{gundavarapu2019sub}%
  \BibitemOpen
  \bibfield  {author} {\bibinfo {author} {\bibfnamefont {S.}~\bibnamefont
  {Gundavarapu}}, \bibinfo {author} {\bibfnamefont {G.~M.}\ \bibnamefont
  {Brodnik}}, \bibinfo {author} {\bibfnamefont {M.}~\bibnamefont {Puckett}},
  \bibinfo {author} {\bibfnamefont {T.}~\bibnamefont {Huffman}}, \bibinfo
  {author} {\bibfnamefont {D.}~\bibnamefont {Bose}}, \bibinfo {author}
  {\bibfnamefont {R.}~\bibnamefont {Behunin}}, \bibinfo {author} {\bibfnamefont
  {J.}~\bibnamefont {Wu}}, \bibinfo {author} {\bibfnamefont {T.}~\bibnamefont
  {Qiu}}, \bibinfo {author} {\bibfnamefont {C.}~\bibnamefont {Pinho}}, \bibinfo
  {author} {\bibfnamefont {N.}~\bibnamefont {Chauhan}},  \emph {et~al.},\
  }\bibfield  {title} {\enquote {\bibinfo {title} {Sub-hertz fundamental
  linewidth photonic integrated brillouin laser},}\ }\href@noop {} {\bibfield
  {journal} {\bibinfo  {journal} {Nature Photonics}\ }\textbf {\bibinfo
  {volume} {13}},\ \bibinfo {pages} {60--67} (\bibinfo {year}
  {2019})}\BibitemShut {NoStop}%
\bibitem [{\citenamefont {Li}, \citenamefont {Lee},\ and\ \citenamefont
  {Vahala}(2013)}]{li2013microwave}%
  \BibitemOpen
  \bibfield  {author} {\bibinfo {author} {\bibfnamefont {J.}~\bibnamefont
  {Li}}, \bibinfo {author} {\bibfnamefont {H.}~\bibnamefont {Lee}}, \ and\
  \bibinfo {author} {\bibfnamefont {K.~J.}\ \bibnamefont {Vahala}},\ }\bibfield
   {title} {\enquote {\bibinfo {title} {Microwave synthesizer using an on-chip
  brillouin oscillator},}\ }\href@noop {} {\bibfield  {journal} {\bibinfo
  {journal} {Nature communications}\ }\textbf {\bibinfo {volume} {4}},\
  \bibinfo {pages} {1--7} (\bibinfo {year} {2013})}\BibitemShut {NoStop}%
\bibitem [{\citenamefont {Zhu}\ \emph {et~al.}(2019)\citenamefont {Zhu},
  \citenamefont {Zohrabi}, \citenamefont {Bae}, \citenamefont {Horning},
  \citenamefont {Grayson}, \citenamefont {Park},\ and\ \citenamefont
  {Gopinath}}]{zhu2019nonlinear}%
  \BibitemOpen
  \bibfield  {author} {\bibinfo {author} {\bibfnamefont {J.}~\bibnamefont
  {Zhu}}, \bibinfo {author} {\bibfnamefont {M.}~\bibnamefont {Zohrabi}},
  \bibinfo {author} {\bibfnamefont {K.}~\bibnamefont {Bae}}, \bibinfo {author}
  {\bibfnamefont {T.~M.}\ \bibnamefont {Horning}}, \bibinfo {author}
  {\bibfnamefont {M.~B.}\ \bibnamefont {Grayson}}, \bibinfo {author}
  {\bibfnamefont {W.}~\bibnamefont {Park}}, \ and\ \bibinfo {author}
  {\bibfnamefont {J.~T.}\ \bibnamefont {Gopinath}},\ }\bibfield  {title}
  {\enquote {\bibinfo {title} {Nonlinear characterization of silica and
  chalcogenide microresonators},}\ }\href@noop {} {\bibfield  {journal}
  {\bibinfo  {journal} {Optica}\ }\textbf {\bibinfo {volume} {6}},\ \bibinfo
  {pages} {716--722} (\bibinfo {year} {2019})}\BibitemShut {NoStop}%
\bibitem [{\citenamefont {Multiphysics}(1998)}]{multiphysics1998introduction}%
  \BibitemOpen
  \bibfield  {author} {\bibinfo {author} {\bibfnamefont {C.}~\bibnamefont
  {Multiphysics}},\ }\bibfield  {title} {\enquote {\bibinfo {title}
  {Introduction to comsol multiphysics{\textregistered}},}\ }\href@noop {}
  {\bibfield  {journal} {\bibinfo  {journal} {COMSOL Multiphysics, Burlington,
  MA, accessed Feb}\ }\textbf {\bibinfo {volume} {9}},\ \bibinfo {pages} {2018}
  (\bibinfo {year} {1998})}\BibitemShut {NoStop}%
\bibitem [{\citenamefont {Debut}, \citenamefont {Randoux},\ and\ \citenamefont
  {Zemmouri}(2000)}]{debut2000linewidth}%
  \BibitemOpen
  \bibfield  {author} {\bibinfo {author} {\bibfnamefont {A.}~\bibnamefont
  {Debut}}, \bibinfo {author} {\bibfnamefont {S.}~\bibnamefont {Randoux}}, \
  and\ \bibinfo {author} {\bibfnamefont {J.}~\bibnamefont {Zemmouri}},\
  }\bibfield  {title} {\enquote {\bibinfo {title} {Linewidth narrowing in
  brillouin lasers: Theoretical analysis},}\ }\href@noop {} {\bibfield
  {journal} {\bibinfo  {journal} {Physical Review A}\ }\textbf {\bibinfo
  {volume} {62}},\ \bibinfo {pages} {023803} (\bibinfo {year}
  {2000})}\BibitemShut {NoStop}%
\bibitem [{\citenamefont {Pant}\ \emph {et~al.}(2011)\citenamefont {Pant},
  \citenamefont {Poulton}, \citenamefont {Choi}, \citenamefont {Mcfarlane},
  \citenamefont {Hile}, \citenamefont {Li}, \citenamefont {Thevenaz},
  \citenamefont {Luther-Davies}, \citenamefont {Madden},\ and\ \citenamefont
  {Eggleton}}]{pant2011chip}%
  \BibitemOpen
  \bibfield  {author} {\bibinfo {author} {\bibfnamefont {R.}~\bibnamefont
  {Pant}}, \bibinfo {author} {\bibfnamefont {C.~G.}\ \bibnamefont {Poulton}},
  \bibinfo {author} {\bibfnamefont {D.-Y.}\ \bibnamefont {Choi}}, \bibinfo
  {author} {\bibfnamefont {H.}~\bibnamefont {Mcfarlane}}, \bibinfo {author}
  {\bibfnamefont {S.}~\bibnamefont {Hile}}, \bibinfo {author} {\bibfnamefont
  {E.}~\bibnamefont {Li}}, \bibinfo {author} {\bibfnamefont {L.}~\bibnamefont
  {Thevenaz}}, \bibinfo {author} {\bibfnamefont {B.}~\bibnamefont
  {Luther-Davies}}, \bibinfo {author} {\bibfnamefont {S.~J.}\ \bibnamefont
  {Madden}}, \ and\ \bibinfo {author} {\bibfnamefont {B.~J.}\ \bibnamefont
  {Eggleton}},\ }\bibfield  {title} {\enquote {\bibinfo {title} {On-chip
  stimulated brillouin scattering},}\ }\href@noop {} {\bibfield  {journal}
  {\bibinfo  {journal} {Optics express}\ }\textbf {\bibinfo {volume} {19}},\
  \bibinfo {pages} {8285--8290} (\bibinfo {year} {2011})}\BibitemShut {NoStop}%
\bibitem [{\citenamefont {Gorodetsky}, \citenamefont {Savchenkov},\ and\
  \citenamefont {Ilchenko}(1996)}]{gorodetsky1996ultimate}%
  \BibitemOpen
  \bibfield  {author} {\bibinfo {author} {\bibfnamefont {M.~L.}\ \bibnamefont
  {Gorodetsky}}, \bibinfo {author} {\bibfnamefont {A.~A.}\ \bibnamefont
  {Savchenkov}}, \ and\ \bibinfo {author} {\bibfnamefont {V.~S.}\ \bibnamefont
  {Ilchenko}},\ }\bibfield  {title} {\enquote {\bibinfo {title} {Ultimate q of
  optical microsphere resonators},}\ }\href@noop {} {\bibfield  {journal}
  {\bibinfo  {journal} {Optics letters}\ }\textbf {\bibinfo {volume} {21}},\
  \bibinfo {pages} {453--455} (\bibinfo {year} {1996})}\BibitemShut {NoStop}%
\bibitem [{\citenamefont {Black}(1957)}]{black1957properties}%
  \BibitemOpen
  \bibfield  {author} {\bibinfo {author} {\bibfnamefont {M.~H.}\ \bibnamefont
  {Black}},\ }\bibfield  {title} {\enquote {\bibinfo {title} {Properties of
  arsenic sulfide glass},}\ }\href@noop {} {\bibfield  {journal} {\bibinfo
  {journal} {Journal of Research of the National Bureau of Standards}\ }\textbf
  {\bibinfo {volume} {59}},\ \bibinfo {pages} {83} (\bibinfo {year}
  {1957})}\BibitemShut {NoStop}%
\bibitem [{\citenamefont {Razdan}\ and\ \citenamefont
  {Van~Baak}(2002)}]{razdan2002demonstrating}%
  \BibitemOpen
  \bibfield  {author} {\bibinfo {author} {\bibfnamefont {K.}~\bibnamefont
  {Razdan}}\ and\ \bibinfo {author} {\bibfnamefont {D.}~\bibnamefont
  {Van~Baak}},\ }\bibfield  {title} {\enquote {\bibinfo {title} {Demonstrating
  optical beat notes through heterodyne experiments},}\ }\href@noop {}
  {\bibfield  {journal} {\bibinfo  {journal} {American Journal of Physics}\
  }\textbf {\bibinfo {volume} {70}},\ \bibinfo {pages} {1061--1067} (\bibinfo
  {year} {2002})}\BibitemShut {NoStop}%
\bibitem [{\citenamefont {{Newport Corporation}}(2020)}]{newportknee}%
  \BibitemOpen
  \bibfield  {author} {\bibinfo {author} {\bibnamefont {{Newport
  Corporation}}},\ }\bibfield  {title} {\enquote {\bibinfo {title} {The
  differences between threshold current calculation methods},}\ }\href@noop {}
  {\  (\bibinfo {year} {2020})}\BibitemShut {NoStop}%
\bibitem [{\citenamefont {Wolff}\ \emph {et~al.}(2017)\citenamefont {Wolff},
  \citenamefont {Stiller}, \citenamefont {Eggleton}, \citenamefont {Steel},\
  and\ \citenamefont {Poulton}}]{wolff2017cascaded}%
  \BibitemOpen
  \bibfield  {author} {\bibinfo {author} {\bibfnamefont {C.}~\bibnamefont
  {Wolff}}, \bibinfo {author} {\bibfnamefont {B.}~\bibnamefont {Stiller}},
  \bibinfo {author} {\bibfnamefont {B.~J.}\ \bibnamefont {Eggleton}}, \bibinfo
  {author} {\bibfnamefont {M.~J.}\ \bibnamefont {Steel}}, \ and\ \bibinfo
  {author} {\bibfnamefont {C.~G.}\ \bibnamefont {Poulton}},\ }\bibfield
  {title} {\enquote {\bibinfo {title} {Cascaded forward brillouin scattering to
  all stokes orders},}\ }\href@noop {} {\bibfield  {journal} {\bibinfo
  {journal} {New Journal of Physics}\ }\textbf {\bibinfo {volume} {19}},\
  \bibinfo {pages} {023021} (\bibinfo {year} {2017})}\BibitemShut {NoStop}%
\bibitem [{\citenamefont {Jin}\ \emph {et~al.}(2017)\citenamefont {Jin},
  \citenamefont {Wang}, \citenamefont {Wang}, \citenamefont {Dong},
  \citenamefont {Li},\ and\ \citenamefont {Wang}}]{jin2017dispersion}%
  \BibitemOpen
  \bibfield  {author} {\bibinfo {author} {\bibfnamefont {X.}~\bibnamefont
  {Jin}}, \bibinfo {author} {\bibfnamefont {J.}~\bibnamefont {Wang}}, \bibinfo
  {author} {\bibfnamefont {M.}~\bibnamefont {Wang}}, \bibinfo {author}
  {\bibfnamefont {Y.}~\bibnamefont {Dong}}, \bibinfo {author} {\bibfnamefont
  {F.}~\bibnamefont {Li}}, \ and\ \bibinfo {author} {\bibfnamefont
  {K.}~\bibnamefont {Wang}},\ }\bibfield  {title} {\enquote {\bibinfo {title}
  {Dispersion engineering of a microsphere via multi-layer coating},}\
  }\href@noop {} {\bibfield  {journal} {\bibinfo  {journal} {Applied Optics}\
  }\textbf {\bibinfo {volume} {56}},\ \bibinfo {pages} {8023--8028} (\bibinfo
  {year} {2017})}\BibitemShut {NoStop}%
\bibitem [{\citenamefont {Zhu}\ \emph {et~al.}(2020)\citenamefont {Zhu},
  \citenamefont {Horning}, \citenamefont {Zohrabi}, \citenamefont {Park},\ and\
  \citenamefont {Gopinath}}]{zhu2020photo}%
  \BibitemOpen
  \bibfield  {author} {\bibinfo {author} {\bibfnamefont {J.}~\bibnamefont
  {Zhu}}, \bibinfo {author} {\bibfnamefont {T.~M.}\ \bibnamefont {Horning}},
  \bibinfo {author} {\bibfnamefont {M.}~\bibnamefont {Zohrabi}}, \bibinfo
  {author} {\bibfnamefont {W.}~\bibnamefont {Park}}, \ and\ \bibinfo {author}
  {\bibfnamefont {J.~T.}\ \bibnamefont {Gopinath}},\ }\bibfield  {title}
  {\enquote {\bibinfo {title} {Photo-induced writing and erasing of gratings in
  as 2 s 3 chalcogenide microresonators},}\ }\href@noop {} {\bibfield
  {journal} {\bibinfo  {journal} {Optica}\ }\textbf {\bibinfo {volume} {7}},\
  \bibinfo {pages} {1645--1648} (\bibinfo {year} {2020})}\BibitemShut {NoStop}%
\bibitem [{\citenamefont {Kang}\ \emph {et~al.}(2017)\citenamefont {Kang},
  \citenamefont {Krogstad}, \citenamefont {Grayson}, \citenamefont {Kim},
  \citenamefont {Lee}, \citenamefont {Gopinath},\ and\ \citenamefont
  {Park}}]{kang2017high}%
  \BibitemOpen
  \bibfield  {author} {\bibinfo {author} {\bibfnamefont {G.}~\bibnamefont
  {Kang}}, \bibinfo {author} {\bibfnamefont {M.~R.}\ \bibnamefont {Krogstad}},
  \bibinfo {author} {\bibfnamefont {M.}~\bibnamefont {Grayson}}, \bibinfo
  {author} {\bibfnamefont {D.-G.}\ \bibnamefont {Kim}}, \bibinfo {author}
  {\bibfnamefont {H.}~\bibnamefont {Lee}}, \bibinfo {author} {\bibfnamefont
  {J.~T.}\ \bibnamefont {Gopinath}}, \ and\ \bibinfo {author} {\bibfnamefont
  {W.}~\bibnamefont {Park}},\ }\bibfield  {title} {\enquote {\bibinfo {title}
  {High quality chalcogenide-silica hybrid wedge resonator},}\ }\href@noop {}
  {\bibfield  {journal} {\bibinfo  {journal} {Optics Express}\ }\textbf
  {\bibinfo {volume} {25}},\ \bibinfo {pages} {15581--15589} (\bibinfo {year}
  {2017})}\BibitemShut {NoStop}%
\bibitem [{\citenamefont {Kittlaus}, \citenamefont {Shin},\ and\ \citenamefont
  {Rakich}(2016)}]{kittlaus2016large}%
  \BibitemOpen
  \bibfield  {author} {\bibinfo {author} {\bibfnamefont {E.~A.}\ \bibnamefont
  {Kittlaus}}, \bibinfo {author} {\bibfnamefont {H.}~\bibnamefont {Shin}}, \
  and\ \bibinfo {author} {\bibfnamefont {P.~T.}\ \bibnamefont {Rakich}},\
  }\bibfield  {title} {\enquote {\bibinfo {title} {Large brillouin
  amplification in silicon},}\ }\href@noop {} {\bibfield  {journal} {\bibinfo
  {journal} {Nature Photonics}\ }\textbf {\bibinfo {volume} {10}},\ \bibinfo
  {pages} {463--467} (\bibinfo {year} {2016})}\BibitemShut {NoStop}%
\bibitem [{\citenamefont {Yu}\ and\ \citenamefont {Sun}(2018)}]{yu2018giant}%
  \BibitemOpen
  \bibfield  {author} {\bibinfo {author} {\bibfnamefont {Z.}~\bibnamefont
  {Yu}}\ and\ \bibinfo {author} {\bibfnamefont {X.}~\bibnamefont {Sun}},\
  }\bibfield  {title} {\enquote {\bibinfo {title} {Giant enhancement of
  stimulated brillouin scattering with engineered phoxonic crystal
  waveguides},}\ }\href@noop {} {\bibfield  {journal} {\bibinfo  {journal}
  {Optics Express}\ }\textbf {\bibinfo {volume} {26}},\ \bibinfo {pages}
  {1255--1267} (\bibinfo {year} {2018})}\BibitemShut {NoStop}%
\end{thebibliography}%

\end{document}


\preprint{AIP/123-QED}

\title{Cascaded forward Brillouin lasing in a chalcogenide whispering gallery mode microresonator - Supplementary Material}
\author{Thariq Shanavas}
 \affiliation{Department of Physics, University of Colorado, Boulder, CO 80309, USA}
\author{Michael Grayson}
\affiliation{Department of Electrical, Computer and Energy Engineering, \\University of Colorado, Boulder, CO 80309, USA}
\author{Bo Xu}
\affiliation{Department of Physics, University of Colorado, Boulder, CO 80309, USA}
\author{Mo Zohrabi}
\affiliation{Department of Electrical, Computer and Energy Engineering, \\University of Colorado, Boulder, CO 80309, USA}
\author{Wounjhang Park}
\affiliation{Department of Electrical, Computer and Energy Engineering, \\University of Colorado, Boulder, CO 80309, USA}
\affiliation{Materials Science Engineering Program, \\University of Colorado, Boulder, CO 80309, USA}
\author{Juliet T. Gopinath}
\affiliation{Department of Physics, University of Colorado, Boulder, CO 80309, USA}
\affiliation{Department of Electrical, Computer and Energy Engineering, \\University of Colorado, Boulder, CO 80309, USA}
\affiliation{Materials Science Engineering Program, \\University of Colorado, Boulder, CO 80309, USA}
\email{julietg@colorado.edu}
\date{\today}

\begin{abstract}
This document contains supplementary material for "Cascaded forward Brillouin lasing in a chalcogenide whispering gallery mode microresonator". Section \ref{antiStokes} contains information on the heterodyne measurement of anti-Stokes beams generated by forward Brillouin scattering in a chalcogenide microresonator.
\end{abstract}

\maketitle

\section{Anti-Stokes beams generated via cascaded forward Brillouin scattering}
\label{antiStokes}

\begin{figure}
    \centering
    \includegraphics[width=\textwidth]{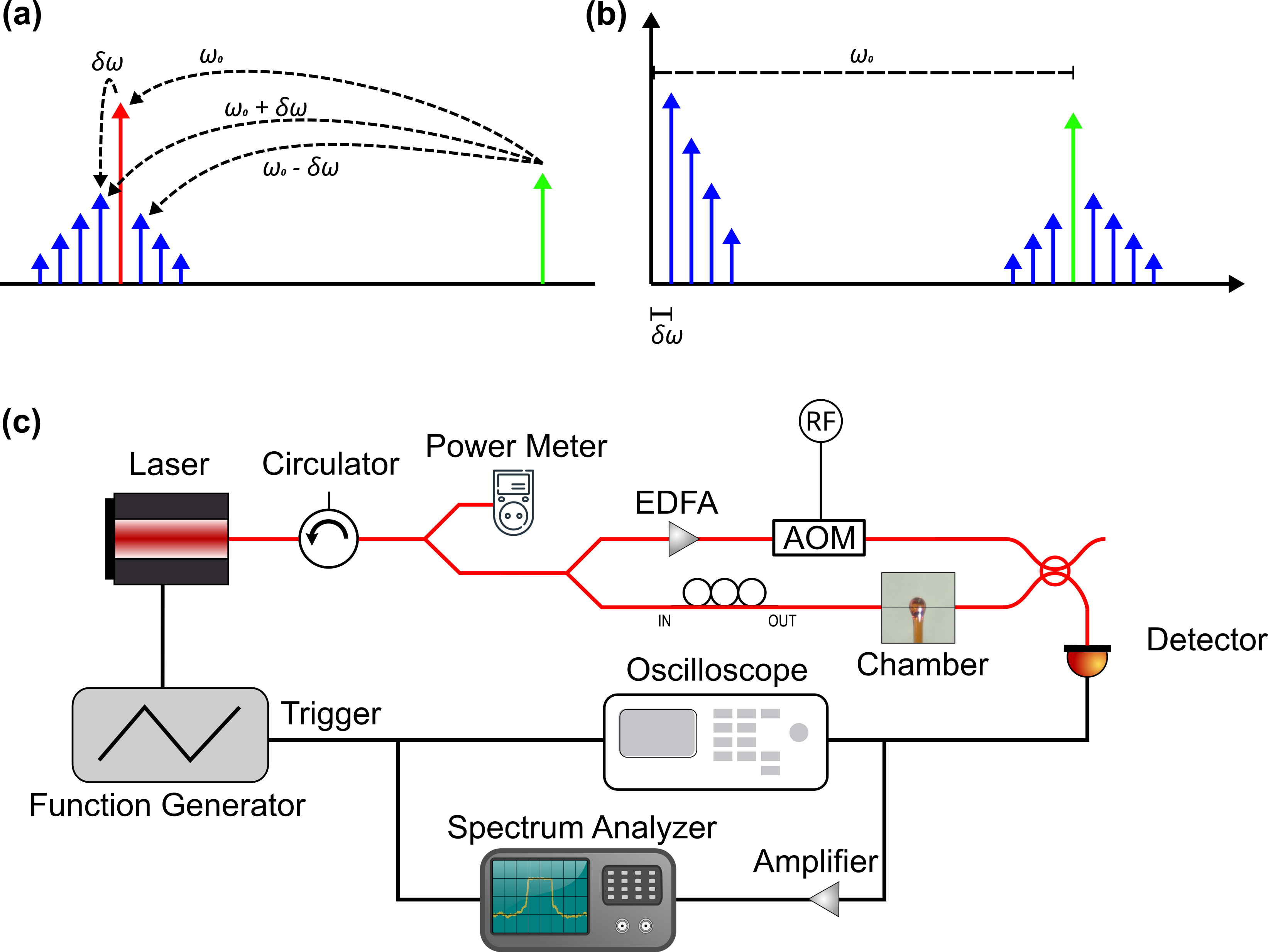}
    \caption{(a) Schematic for optical heterodyne measurement of anti-Stokes beams generated by cascaded forward Brillouin scattering. The acousto-optic modulator (AOM) generates a probe beam with a frequency 200 MHz higher than the pump frequency. The beat notes are generated at $\omega_0\pm n \delta\omega$, where $\omega_0$ = 200 MHz is the drive frequency of the AOM and $\delta\omega$ is the Brillouin shift. (b) Beat notes generated between the cascaded Brillouin beams, pump beam and probe beam. The beat notes from the Stokes beams appear at a higher frequency than the AOM drive frequency. (c) Experimental setup for heterodyne measurement of anti-Stokes beams. The tunable diode at 1550 nm was coupled into a 100$\upmu$m microsphere using a tapered silica fiber and tuned into resonance. The AOM generates a heterodyne probe beam, with no modulation applied. To compensate for the relatively large insertion loss of the AOM, an Erbium-doped fiber amplifier (EDFA) was used. A 50-50 coupler was used to mix the probe beam with the signal from the resonator. The beatnotes are detected using a detector of bandwidth 1.2 GHz and measured using an RF spectrum analyzer.}
    \label{fig:AOM_heterodyne}
\end{figure}

\begin{figure}
    \centering
    \includegraphics[width=\textwidth]{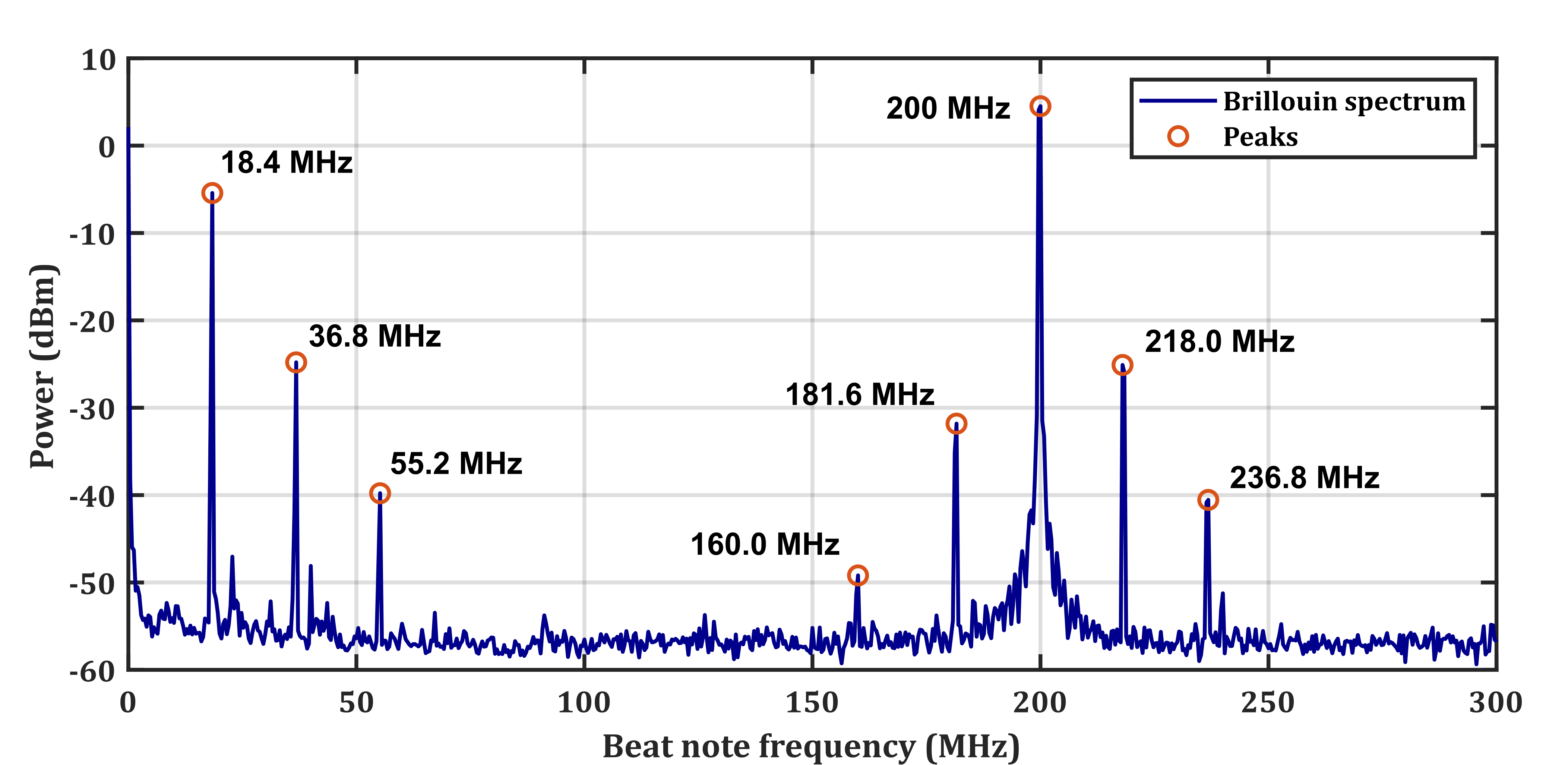}
    \caption{Cascaded FSBS spectrum in a 100 $\upmu$m sphere with unloaded quality factor $2\times 10^6$ measured using a heterodyne probe beam. The anti-Stokes and Stokes beams generate beat notes to the left and right of the AOM drive frequency at 200 MHz, respectively. In the low frequency end, the beat notes between the anti-Stokes and the pump overlap with the beat notes between the pump and the Stokes beams. Near the AOM drive frequency, the anti-Stokes and Stokes beams are mapped to unique frequencies in the RF domain. We note that the Stokes beams are stronger than the anti-Stokes beams.}
    \label{fig:AOM_heterodyne_data}
\end{figure}
Previous theoretical work has established that anti-Stokes beams will be generated along with Stokes beams in a waveguide beyond the threshold for cascaded forward Brillouin scattering\cite{wolff2017cascaded}. The analysis also applies to resonators that are simultaneously resonant for the pump, Stokes, anti-Stokes and acoustic waves. However, a direct measurement of the beat note between the pump and the anti-Stokes beams does not allow an independent measurement of the anti-Stokes beams. This is because the beat note between the pump and the anti-Stokes beam has the same frequency as the beat note between the pump and the Stokes beam and the latter eclipses the signal from the anti-Stokes beam, as Stokes beams are stronger than anti-Stokes beams. It is also not easy to resolve the Stokes and anti-Stokes beams using an optical spectrum analyzer since they are separated by less than 100 MHz. Therefore, it is necessary to perform a heterodyne measurement with a probe laser at a frequency different from the pump beam to ensure the Stokes and anti-Stokes beams are mapped to unique beat frequencies. The schematic for this measurement is shown in Fig. \ref{fig:AOM_heterodyne}. The quality factor of the microresonator used for this measurement was $2\times 10^6$ and had a diameter of 100$\upmu$m. The resonator was undercoupled in this measurement. The full width at half-maximum is about 100 MHz, so we do not expect a large number of cascaded Stokes beams.

Fig. \ref{fig:AOM_heterodyne_data} shows the beat notes from the cascaded FSBS in the microsphere. The anti-Stokes and Stokes beams generate beat notes to the left and right of the AOM drive frequency at 200 MHz, respectively. In the low frequency end, the beat notes between the anti-Stokes and the pump overlap with the beat notes between the pump and the Stokes beams. Near the AOM drive frequency, the anti-Stokes and Stokes beams are mapped to unique frequencies in the RF domain. The Stokes beams are observed to be stronger than the anti-Stokes beams.

\bibliography{aipsamp}